\documentclass[12pt,preprint]{aastex}
\begin{document}

\parindent=1.0cm

\title {SHAKEN, NOT STIRRED: THE DISRUPTED DISK OF THE STARBURST GALAXY NGC 253 \altaffilmark{1}}

\author{T. J. Davidge}

\affil{Herzberg Institute of Astrophysics,
\\National Research Council of Canada, 5071 West Saanich Road,
\\Victoria, BC Canada V9E 2E7\\ {\it email: tim.davidge@nrc.ca}}

\altaffiltext{1}{Based on observations obtained with WIRCam, a joint project of
the CFHT, Taiwan, Korea, Canada and France, at the Canada-France-Hawaii Telescope,
which is operated by the National Research Council of Canada, the Institut National
des Sciences de l'Univers of the Centre National de la Recherche Scientifique of
France, and the University of Hawaii.}

\begin{abstract}

	Near-infrared images obtained with WIRCam on the Canada-France-Hawaii 
Telescope are used to investigate the recent history 
of the nearby Sculptor Group spiral NGC 253, which is one of the nearest starburst 
galaxies. Bright AGB stars are traced out to projected distances of 
$\sim 22$ -- 26 kpc ($\sim 13 - 15$ disk scale lengths) along the major axis. 
The distribution of stars in the disk is lop-sided, in the sense that the projected 
density of AGB stars in the north east portion of the disk between 10 
and 20 kpc from the galaxy center is $\sim 0.5$ dex higher than on the opposite 
side of the galaxy. A large population of red supergiants is also found in the north 
east portion of the disk and, with the exception of the central 2 kpc, 
this area appears to have been the site of the highest levels of star-forming 
activity in the galaxy during the past $\sim 0.1$ Gyr. 
It is argued that such high levels of localized star 
formation may have produced a fountain that ejected material from the disk, and 
the extraplanar HI detected by Boomsma et al. (2005) may be one 
manifestation of such activity. Diffuse stellar structures are found in the periphery 
of the disk, and the most prominent of these is to the south and east of the 
galaxy. Bright AGB stars, including cool C stars that are identified based 
on their $J-K$ colors, are detected out to 15 kpc above the disk plane, and these are part 
of a diffusely distributed, flattened extraplanar component. Comparisons between observed 
and model luminosity functions suggest that the extraplanar regions contain stars that 
formed throughout much of the age of the Universe. Additional evidence 
of a diffuse, extraplanar stellar component that contains 
moderately young stars comes from archival GALEX images. 
It is suggested that the disk of NGC 253 was disrupted by 
a tidal encounter with a now defunct companion. This encounter 
introduced asymmetries that remain to this day, and the projected distribution of stars in 
and around NGC 253 suggests that the companion had an orbit that was 
prograde and moderately inclined to the NGC 253 disk. 
The star-forming history of the extraplanar stars suggests that they either
originated in the NGC 253 disk, or in a gas-rich companion. 
In the latter case the companion must have had an initial M$_B < -15$ 
in order to produce the more-or-less continuous star-forming history that is suggested 
by the stellar content. The ages of the youngest extraplanar stars suggests that 
the event that produced the extraplanar population, and 
presumably induced the starburst, occured within the past $\sim 0.2$ Gyr.

\end{abstract}


\section{INTRODUCTION}

	Nearby galaxies are unique laboratories for investigating the processes 
that drive galaxy evolution. The information obtained from 
the photometric and spectroscopic analyses of individual stars can be employed to probe 
the history of nearby galaxies to a level of detail that is not possible for 
distant systems, where stellar content studies are limited to information gleaned 
from the analysis of integrated light. Star counts also provide a means of tracing 
structures to lower densities than is possible with integrated light, thereby allowing 
diffuse tidal debris trails and low surface brightness companions, both of which provide 
clues into the past history of a galaxy, to be identified and characterized.

	The majority of large galaxies within $\sim 10$ Mpc 
are disk-dominated, and studies of their resolved stellar contents reveal 
considerable diversity in recent star-forming histories. Galaxies such as M31 (Williams 
2002) and NGC 5102 (Davidge 2008a; 2010) have depressed star formation rates (SFRs) 
at the present day when compared with $\sim 1$ Gyr in the past. In contrast, a number of 
nearby galaxies, such as M82, NGC 5253, NGC 253, and M83, have elevated levels of 
central star-forming activity. That a number of nearby galaxies are experiencing elevated 
levels of star-forming activity indicates that this is not a rare phenomenon 
in the local Universe, and starbursts account for at least 20\% of 
all recent local star formation (Brinchmann et al. 2004; Lee et al. 2009).

	There is a compelling association between elevated SFRs and galaxy-galaxy 
interactions (e.g. Larson \& Tinsley 1978; Iono et al. 
2004; Kewley et al. 2006). This suggests that tidal effects 
are the primary cause of starbursts, at least among massive galaxies, 
although interactions with intergalactic material can also trigger elevated star-forming 
activity (e.g. Kapferer et al. 2009). It is thus worth noting that some of the nearest 
starburst or post-starburst galaxies do not have obvious companions or are not located 
in environments that are thought to contain a dense intergalactic medium, leaving 
no obvious trigger for the starburst activity. Two such examples are NGC 5102 
and NGC 253. The former galaxy was investigated by Davidge (2008c), while the latter is the 
subject of this paper.

	NGC 253 is the largest member of the filamentary collection of galaxies 
that is classically referred to as the Sculptor group. With a crossing time of 
6 -- 7 Gyr (Karachensev 2005), the group is not dynamically relaxed, 
and interactions between its dominant members are likely rare. Still, NGC 253 has 
not evolved in isolation. The spin of the NGC 253 disk is consistent with torquing 
from NGC 247 (Whiting 1999), which -- at a projected distance of $\sim 350$ kpc from 
NGC 253 -- is its closest large companion. Extragalactic HI, 
which is a possible signature of an interaction that affected the disk of one or 
both galaxies, has not been detected in the vicinity of NGC 253 and NGC 247 (Putman et 
al. 2003), although neutral hydrogen may be ionized by background radiation 
at large distances from galaxies (e.g. Westmeier, Bruns, \& Kerp 2008). 

	Despite its isolated location, kinematic studies suggest that NGC 253 was involved 
in a recent merger. Anantharamaiah \& Goss (1996) and Prada et al. (1998) find evidence of 
two kinematically distinct systems near the center of NGC 253. One system has an axis of 
rotation that differs from that of the galaxy disk, while the other 
appears to be counter-rotating with respect to the galaxy disk.
Das et al. (2001) model the kinematics of material near the center of NGC 253 and 
detect twisting in the velocity field $\sim 30$ arcsec from the nucleus. 
All three studies interpret their results as kinematic signatures of 
a merger. Das et al. (2001) estimate that the merger may have involved the accretion of 
$10^6$ M$_{\odot}$ of material $10^7$ years in the past, and they 
point out that this may be the remnants of what was initially a much larger object.

	Much of the literature on NGC 253 deals with 
the central regions of the galaxy and the superwind. 
Rieke \& Low (1975) first noted that the photometric properties of the central regions of 
the galaxy were consistent with a transient level of elevated star formation, and the 
presence of a large central young population was confirmed with the discovery that the 
near-infrared light is dominated by red supergiants (RSGs; Rieke et al. 1980). Rieke et 
al. (1988) suggest that NGC 253 may be in an earlier phase of starburst activity than M82, 
thereby making it a potentially important object for understanding the early phases of 
starburst evolution. 

	The central velocity field defined by CO emission in NGC 253 is indicative of 
a bar, which is likely a conduit for material that feeds the 
central starburst (Paglione et al. 2004). The present-day 
SFR of NGC 253 is 5 M$_{\odot}$ year$^{-1}$ (Melo et al. 2002), and the SNe rate 
is 0.2 year$^{-1}$ (Pietsch et al. 2001). Engelbracht et al. (1998) estimate that 
this level of nuclear star formation has been on-going for 20 -- 30 Myr, although the
SFR may have declined during the past 5 Myr. Starburst activity 
has almost certainly occured for a much longer time than calculated by Engelbracht et al. 
(1998), as star-forming activity in starbursts appears to be variable in both 
time and location (e.g. Forster Schreiber et al. 2003; McQuinn et al. 2009). 

	NGC 253 is interacting with its immediate surroundings. The extraplanar regions 
near the minor axis of NGC 253 are populated by a mix of hot gas, cool gas, and dust, 
and Heckman et al. (1990) estimate that 1.2 M$_{\odot}$ year$^{-1}$ is ejected 
from the central regions. As in the M82 outflow, X-ray and H$\alpha$ emission 
in the NGC 253 outflow occurs in close proximity. The X-ray and H$\alpha$ emission 
likely originates in a boundary environment where the superwind interacts with a dense
cloud of cool material (e.g. McCarthy et al. 1987; Strickland et al. 2002). The cool 
material is presumably tidal in origin, although it could also be remnant natal gas.
There is prominent extraplanar UV emission, and Hoopes et al. (2005) suggest that this is 
light from the starburst that is reflected by dust. Beck et al. (1982) and Malin \& Hadley 
(1997) detect diffuse extraplanar light from NGC 253 at visible wavelengths, and some of 
this light may be emission from the outflow.

	Gas and dust in NGC 253 are centrally concentrated. 
Interstellar material accounts for 7\% of the mass near the center of NGC 253, 
but only 4\% of the mass at larger radii (Scoville et al. 1985).
The radial distribution of molecular material in NGC 253 is a factor of 2 more 
concentrated than in other nearby galaxies (Sorai et al. 2000), and 
there is also a central HI concentration (Combes et al. 1977). 
A large reservoir of molecular material is
found in a star-forming ring that is 2 kpc from the nucleus 
(Sorai et al. 2000), and the spiral structure of NGC 253 
appears to eminate from this structure (Hoopes et al. 1996).

	While most spiral galaxies have HI envelopes that extend well past the optical 
limits of the disk, Puche et al. (1991) and Boomsma et al. (1995) find that the HI disk in 
NGC 253 extends to only $0.8 \times$ the Holmberg radius. In addition, the HI distribution 
in the disk is not symmetric, in the sense that HI in the south west quadrant 
extends to larger radii than in the north east quadrant (Boomsma et al. 2005). 
As noted by Boomsma et al. (2005), the lack of HI emission at large radii may not signal an 
absence of gas in the outer disk; rather, any HI may be ionized if the disk is 
warped at large radii, such that the gas is exposed to radiation 
from the nucleus. Tidal effects could also induce a truncated HI 
distribution, and it is intriguing that the closest large neighbor 
of NGC 253, NGC 247, also has a relatively compact HI distribution (Carignan \& Puche 1990). 

	The elevated levels of star-forming activity in NGC 253 have not been 
restricted to the central regions of the galaxy. 
The NGC 253 disk contains numerous filamentary and bubble-like 
structures that are signatures of a high SFR, and such structures may 
provide outlets for interstellar material to escape from the disk (Sofue et al. 1994). 
In fact, Boomsma et al. (2005) detect diffuse extraplanar HI in the eastern half of 
NGC 253. The HI is detected out to distances of 12 kpc from the disk plane, and the 
extraplanar component accounts for 3\% of the total HI mass of NGC 253. 

	As one of the nearest starburst galaxies, NGC 253 is a prime target for 
stellar content studies, although only a modest number of investigations of its resolved 
stellar content have been published to date. The luminosity 
distribution of x-ray sources in the NGC 253 disk is similar to that in 
M33 and M31, and this is suggestive of a common source population, with the caveat that the 
overlap between the NGC 253 and Local Group LFs is modest (Vogler \& Pietsch 1999). 
Davidge et al. (1991) discuss deep multicolor observations of a field along the 
north east major axis of the disk. A distance modulus of 27 was computed 
based on the properties of the detected RSGs and blue supergiants (BSGs). The photometric 
properties of these stars change with location, suggesting that there are substantive 
spatial variations in extinction across the disk. 

	Karachentsev et al. (2003) discuss deep images of an area of the NGC 
253 disk that is close to the Davidge et al. (1991) field. The CMD in 
Figure 2 of Karachentsev et al. (2003) is dominated by a broad RGB that is 
capped by a richly populated AGB. Karachentsev et al. (2003) compute a distance modulus 
of 27.98 from the RGB-tip, and this distance modulus is adopted for the present study.
There is also a bright main sequence in the Karachentsev et al. CMD, which has 
a width of $\pm 0.3$ mag in $V-I$. This dispersion exemplifies 
the extent of differential reddening in the NGC 253 disk.

	NGC 253 was imaged as part of the ANGST survey (Dalcanton et al. 2009). 
In accordance with the survey guidelines, the fields 
observed in NGC 253 were restricted to the major axis, and cover roughly 
one half of the disk. While there is a single major axis field at moderately 
large distance from the galaxy center, the ANGST data do not 
sample the extraplanar regions of the galaxy. The main sequence in some of the ANGST 
CMDs of NGC 253 has a width of 1 - 1.5 mag in F475W -- F814W, highlighting the influence of 
differential reddening in photometric studies of the galaxy at visible wavelengths. 
In fact, an inspection of the CMDs in Figures 9 -- 
22 of Dalcanton et al. (2009) indicates that the extent of differential extinction 
in NGC 253 exceeds that in other galaxies in the ANGST sample.

	As with the disk, there have been only a modest number of studies 
of resolved stars in the extraplanar regions of NGC 253. In the first deep CCD 
study of resolved stars in NGC 253, Davidge \& Pritchet (1990) observed a field 
along the north west arm of the minor axis. The completeness-corrected CMD shown in 
their Figure 8 includes the upper portions of the RGB, and there is a break near 
$I = 24$ in their CMD that corresponds to the RGB-tip found in subsequent studies. 
Assuming a stellar content like that in the outer regions of M31, a distance modulus of 
26.8 was computed from the brightest star. This places NGC 253 much closer than more 
recent distance estimates (e.g. Karachentsev et al. 2003; Mouhcine et al. 2005) - 
a result that may be explained if this part of NGC 253 contains a young population 
(e.g. Comeron et al. 2001, and this study), rendering the assumption of a stellar content 
like that in the outer regions of M31 invalid. 

	Comeron et al. (2001) detect bright main sequence stars along the 
south east minor axis of NGC 253 out to distances of 15 kpc from the disk plane. Using 
stellar lifetime arguments, Comeron et al. (2001) suggest that the stars that are furthest 
from the disk plane must have formed in the outflow; a similar conclusion was reached 
independently for young stars in the extraplanar regions of M82 (Davidge 2008c). 
Mouhcine (2005) discusses deeper observations of a portion of the area surveyed by 
Comeron et al. (2001). A population of RGB stars was detected, 
from which a distance modulus of 27.6 was deduced based on the RGB-tip brightness. 
With the caveat that young and intermediate age stars 
may bias the distance estimate and skew the shape of the RGB, 
the photometrically derived metallicity distribution function (MDF) of RGB stars in this 
field peaks near [M/H] = --0.7, with a metal-poor tail that extends to --2 (Mouhcine 2006). 
Such an MDF is not characteristic of a classical old, metal-poor halo.
Rather, this MDF is reminsiscent of that seen throughout much of M31 (e.g. Bellazzini 
et al. 2003; Koch et al. 2008; Richardson et al. 2009; Tanaka et al. 2010), which 
is a galaxy that has experienced interactions with its satellites (e.g. Ibata et al. 2007).

	Dust has a major influence on the appearance of NGC 253 (Sofue et al. 1994), and at 
least 90\% of the H$\alpha$ flux from the galaxy is obscured (Hoopes et al. 1996). The dust 
is widespread, and the disk contains 94\% of the total cool dust supply (Melo et al. 2002). 
The emission from cool dust has a scale length that is similar to (Melo et al. 2002) or 
possibly even larger than (Radovich et al 2001) that of the stellar light. 
The high dust content of the NGC 253 disk provides a compelling 
motivation to survey its stellar content at wavelengths longward of 
$1\mu$m, where the impact of extinction is greatly reduced with respect to 
visible wavelengths. The impact of line blanketing on the 
photometric properties of the reddest stars, such as those 
that are evolving near the asymptotic giant branch (AGB) tip, is also greatly 
reduced in the infrared when compared with the visible.
This enhances the utility of these objects as probes of stellar content, 
as it increases the contrast between the brightest red stars and the 
underlying blue body of the disk, thereby reducing the affect of crowding. 

	In the present paper, near-infrared observations of three $20 \times 20$ arcmin 
fields, recorded with the CFHT WIRCam, are used to map the projected distribution 
of the brightest red stars in the disk and extraplanar regions of NGC 253. The data, 
the basic reduction procedures, and the photometric measurements 
are described in \S 2, while the spatial distribution of various stellar 
types are described in \S 3. The properties of the brightest 
red stars in the disk plane and extrapalanar regions are examined in \S 4 and \S 5, 
respectively. A summary and discussion of the results follows in \S 6. The discovery of an 
extended halo around NGC 253 at ultraviolet (UV) wavelengths, which is distinct from 
the starburst outflow and appears to be due to an intermediate age extraplanar 
component (\S 5), is reported in the Appendix.

\section{OBSERVATIONS, REDUCTIONS, \& PHOTOMETRIC MEASUREMENTS}

\subsection{The Observations and Their Reduction}

	The data were recorded with WIRCam (Puget et al. 2004) on the Canada-France-Hawaii 
Telescope (CFHT) as part of the 2007B observing queue. The WIRCam detector 
is a mosaic of four $2048 \times 2048$ HgCdTe arrays that are deployed in a $2 
\times 2$ format. There is a 45 arcsec gap between detectors, and each 
pixel subtends $0.3 \times 0.3$ arcsec$^2$. Thus, a single exposure 
covers $21 \times 21$ arcmin$^2$.

	Three fields, that sample the major axis of NGC 253 at both ends of the disk 
(Fields 1 and 2) and the south east segment of the minor axis (Field 3), were observed 
through $J$ and $Ks$ filters. The central co-ordinates of these fields are listed in 
Table 1. The locations of the fields on the sky are indicated in Figure 1.
The positioning of Fields 1 and 2 avoids the innermost regions of the 
galaxy, where resolving stars is problematic; hence, the main body of NGC 253 is not 
imaged in its entirety. 

	A five point dither pattern, defining a $5 \times 5$ arcsec square on the sky with a 
central point, was employed to facilitate the identification and the suppression of bad 
pixels and cosmic rays. A dither pattern with a substantially larger throw could have been 
used to fill the gaps between detectors. However, it was decided not to dither over 
large angular scales since a goal of this study is to assess the 
distribution of stars in the outer regions of NGC 253, and large dither offsets 
would lower the maximum distance that would be sampled with full data coverage. 
While the gaps between detectors are cosmetically unappealing, they do not hinder 
the ability to chart large scale trends in the stellar distribution.

	Four 20 sec exposures were obtained per filter at each dither position. The dither 
pattern was repeated 7 times in $J$ and 14 times in $Ks$ for each field. 
The total exposure time per field is thus 2800 sec in $J$ and 5600 sec in $Ks$. 

	The data were reduced using standard techniques for near-infrared imaging. 
The initial processing, which was done with the I'IWI pipeline at the CFHT, 
consisted of dark subtraction and flat-fielding. A calibration frame to remove interference 
fringes and the thermal signatures of warm objects along the light path was 
constructed by combining the I'IWI-processed images of the three fields. 
A mean sky level was subtracted from each image prior to combination, 
and the portions of Fields 1 and 2 that have high stellar density 
were masked to prevent introducing artifacts into the final calibration frame. The
calibration frames constructed in this manner were subtracted from the flat-fielded data, 
and the results were spatially registered and stacked to obtain the final datasets for 
photometric analysis. Stars in the final images have FWHM $\sim 0.9$ arcsec.

\subsection{Photometric Measurements}

	The photometric measurements were made with the point spread function 
(PSF) fitting routine ALLSTAR (Stetson \& Harris 1988). The source catalogs, initial 
magnitudes, and PSFs that are used by ALLSTAR were obtained from 
routines in DAOPHOT (Stetson 1987). The criteria for selecting PSF stars were 
a stellar appearance, a brightness that was well above the faint limit of the 
data, and a lack of bright neighbors. Each PSF is the result of combining roughly 50 
stars. Faint sources close to the PSF stars were removed iteratively. 
The photometry was calibrated using standard star observations that were recorded 
over the same time period as the NGC 253 observations, and the results were checked 
against measurements in the 2MASS point source catalogue. The mean difference between 
the WIRCam and 2MASS measurements is 0.008 magnitudes in $K$, with a standard deviation of 
$\pm 0.044$ magnitudes.

	The photometric catalogue was filtered to remove objects with 
PSF-fit errors, $\epsilon$, $\geq 0.3$ magnitude. This 
defines the faint limit of the photometry, rejecting objects for which 
photometry is problematic owing to a low S/N ratio. 
Objects that departed from the dominant relation between $\epsilon$ and magnitude 
were also removed, and these tend to be sources that are either moderately extended, 
in very crowd environments, associated with diffraction spikes, or 
cosmetic defects. Finally, objects with a DAOPHOT ROUND parameter that 
is markedly larger than the norm, which tend to be either galaxies or blended stars, 
were also removed from the photometric catalogue.

	The majority of sources that were removed from the photometric catalogue were 
identified with the two $\epsilon$ criteria. An example of the relation used to cull 
objects based on $\epsilon$, and the type of objects that are 
removed or retained, is shown in Figure 2 for sources in Field 3. 
The dashed line in Figure 2 is the curve that was used to reject sources; 
objects that fall above this curve were deleted from the photometric catalogue.

	The two highlighted objects with $\epsilon$ values that fall above the dashed line 
in Figure 2 are clearly galaxies, while the two highlighted objects 
that fall below this line have a stellar appearance. The culling 
function that was applied is not overly zealous, and a more draconian cut could have been 
applied that runs much closer to the stellar locus near $\epsilon = 0.01 - 0.02$. However, 
it was decided to err on the side of caution, as there is not a clear-cut division between 
stars and galaxies. Extended objects that remain after this initial culling are accounted 
for in a statistic fashion, using source counts made at large distances from NGC 253.

	Artificial star experiments were run to assess sample completeness and estimate the 
uncertainties in the photometeric measurements that are due to photon noise and crowding. 
Artificial stars were assigned $J-K = 1.3$, which is the approximate color of the bright 
end of the AGB in NGC 253. Like actual stars in NGC 253, an artificial star was only 
considered to be detected if it was recovered in both filters. The artificial stars were 
subject to the $\epsilon$ and ROUND culling criteria described earlier in this \S.

	The uncertainties in the photometric measurements increase rapidly near 
the magnitude at which the sample is 50\% complete. The 50\% completeness 
magnitude varies across the field, in the sense that it 
becomes brighter towards progressively more crowded environments. The stellar catalogues 
are 50\% complete near $K \sim 20.5$ in the disk 10 kpc from the galaxy 
center, and near $K \sim 21.5$ in the least crowded portions of all three fields. 
These magnitudes are consistent with the general appearance of the 
luminosity functions (LFs) discussed in Sections 4 and 5.

\section{THE STELLAR DISTRIBUTION: AN OVERVIEW}

	Many nearby galaxies that have superficially mundane morphologies, are actually 
embedded in highly complex stellar distributions, with sub-structures that are relics of 
past interactions. As the nearest large external spiral galaxy, M31 has been the 
target of numerous studies (e.g. Tanaka et al. 2010; Ibata et al. 2007, 
and references therein). During the past few years efforts to identify and 
characterize low surface brightness structures have been extended to more 
distant galaxies, including M81 (Davidge 2008d; Mouhcine \& Ibata 2009), 
M82 (Davidge 2008c), NGC 5102 (Davidge 2010), NGC 891 (Mouhcine et al. 2010), NGC 5907 
(Martinez-Delgado et al. 2008), and NGC 4013 (Martinez-Delgado et al. 2009). These studies 
reveal that tidal structures are a common feature among nearby 
spiral galaxies, as might be anticipated given the angular momentum content 
of their disks (e.g. Hammer et al. 2007).

	There are indications that NGC 253 may also have extraplanar stellar
sub-structures. Beck et al. (1982) and Malin \& Hadley (1997) detect diffuse extraplanar 
structures in deep photographic images of NGC 253. While the nature of these 
structures is not clear, they should be readily apparent in images recorded with 
digital detectors. Moreover, the brightest stars in these structures 
should also be detectable during good seeing conditions. In this section, 
the overall stellar distribution in NGC 253 is discussed, with emphasis on identifying 
features and trends that will be discussed in greater detail in subsequent sections.

\subsection{Structures in the Stellar Distribution}

	Samples of M giants, C stars, and RSGs have been defined 
based on location in the $(K, J-K)$ CMD. The criteria used here are indicated 
in the right hand panel of Figure 6 (\S 4), and Figure 3 shows the spatial distribution of 
each stellar type. A large fraction of the M giant and C star samples at large 
distances from the center of NGC 253 are actually background galaxies. 
M giants far outnumber C stars and RSGs, and the distribution of 
M giants in Figure 3 is shown with an expanded scale to facilitate the 
identification of subtle features. 

	Two distinct cross patterns, made up of columns and rows in which stars are abscent, 
are evident in the upper left hand detector of Field 1. These are centered on bright 
stars that were targeted by the WIRCam image stabilization unit, which repeatedly reads the 
signal from a small region centered on selected stars to fine-tune guiding. While the use of 
bright stars as guide beacons negates the use of the columns and rows 
that converge on them for science, the rest of the frame is not affected.

	While all of the stars considered here are in short-lived stages of 
evolution, and hence are relatively rare, several structures can still be identified 
in the stellar distributions. The disk of NGC 253 is the most 
prominent structure in the distribution of all three stellar types. 
The cleanest disk distribution is defined by RSGs; not only are the ages of 
RSGs such that they are largely confined to the disk plane, but contamination from 
foreground stars and background galaxies is also modest in the part of the 
CMDs that contains RSGs. There is a clear concentration of RSGs in the north east 
part of the disk, indicating that this area had a higher SFR than elsewhere in the disk 
during the past few tens of Myr.

	The bright M giants in the disk define a ring-like structure with 
dimensions that are similar to the RSG distribution. The central 
`hole' in the M giant distribution is due to crowding in 
the dense inner disk, which hinders the detection of individual stars. 
Crowding is less of an issue for RSGs in this part of NGC 253, as these stars are 
significantly brighter than M giants. 

	While RSGs are confined to a tight disk-like distribution, M giants are found over 
a larger area. The outer boundary of the M giant distribution is relatively diffuse when 
compared with the RSG distribution, and the M giants also define structures that 
may be signatures of an interaction with a companion. The most prominent such structure is 
the spur in the M giant distribution that extends to the south east of 
the disk in Field 1. This feature is also evident in the deep images discussed 
by Beck et al. (1982) and Malin \& Hadley (1997). This is an area of H$\alpha$ emission, 
but there is no obvious UV emission, suggesting that there is not a large concentration 
of young stars (Hoopes et al. 2005). Figure 2 of Boomsma et al. (2005) shows 
a nose in the outer contour of the HI distribution in this part of NGC 253, 
suggesting that neutral gas may also be present. This spur is not evident in the C star and 
RSG distributions, although this may be due to small number statistics. Structures similar 
to this can result from the disruption of a satellite system (e.g. Figure 2 of Fardal et 
al. 2007). If NGC 253 was involved in such an event then deeper imaging surveys may find 
a stream that defines the debris trail left by the companion galaxy. 

	Bright M giants can be traced to larger distances along the major axis than RSGs, 
and this suggests that the most recent episodes of star formation have been confined 
to small and intermediate radii in the disk. This is contrary to 
what is seen in the outer regions of many other nearby galaxies, such as M83 (Davidge 2010), 
where the luminosity-weighted age drops with increasing radius. An absence of 
gas, and hence young stars, in the outer disk may be a consequence of tidal interactions, 
as gas is funnelled towards the center of the dominant galaxy and stars are scattered 
outwards (e.g. Hopkins et al. 2009). 

	The distribution of M giants in the disk of NGC 253 is lop-sided, in the sense 
that the density of M giants in the outer disk is higher in the north west quadrant than in 
the south east quadrant. The lop-sided distribution of young stars is 
also evident in the distribution of UV light (\S 6.1.2).
Zaritsky \& Rix (1997) investigate the incidence of such asymmetries in nearby galaxies, 
and find that they are not rare. They also find that lop-sided galaxies tend to have 
higher SFRs than symmetric galaxies, and attribute the morphological asymmetry 
to the accretion of a companion. 

	With the caveat that there are differences in the number of sources, 
the C star and M giant distributions share many similarities. One exception is 
a possible concentration of C stars near the minor axis of NGC 253 
that is much less obvious in the M giant distribution. The number density of C stars in this 
part of NGC 253 exceeds that near the edges of the WIRCam survey, 
where contaminating objects presumably dominate, at roughly the $3\sigma$ level.

	While the presence of stars near the minor axis may suggest an 
association with the outflow from the central starburst, 
such a connection is far from ironclad. At present, outflow emission on the south east 
side of the galaxy is offset from the minor axis (Figure 1 of Hoopes et al. 2005). 
Perhaps the most compelling reason not to associate the C stars with the outflow is that 
C stars have ages that are in excess of 0.2 Gyr. 
Based on dynamical arguments and the physical extent of the region affected by the outflow, 
Pietsch et al. (2000) estimate that the outflow has an age of at least 25 Myr, which is 
one to two orders of magnitude shorter than the time frame over which C stars form. 

	Another challenge is to explain why the C stars have 
not dissipated away from the minor axis into the extraplanar field. 
A long-lived sequence of C stars may result if -- for example -- they are 
part of a coherent stream on a circumpolar orbit. If so, then C stars will 
also be seen along the north west segment of the minor axis, which was not 
observed with WIRCam. While a highly inclined orbital geometry of 
the supposed companion might seem contrived, the kinematics of 
material near the center of NGC 253 is consistent with the accretion of an object that was 
on such an orbit (Anantharamaiah \& Goss 1996; Prada et al. 1998).

\subsection{Distinguishing Between Disk and Extraplanar Stars}

	With the exception of foreground stars and background 
galaxies, the WIRCam fields contain a mix of sources that belong to the disk 
and the extraplanar regions of NGC 253. In the present work, 
the sorting of the sources into disk and extraplanar components is done using 
distance from the principle axis of the galaxy, which is a reasonable starting 
point for investigating large-scale spatial trends. Such a classification is possible given 
that the disk of NGC 253 is viewed almost edge-on.

	To be sure, sorting objects into broad `disk' and `extraplanar' components 
is simplistic, and does not address the presence of physically distinct structures. 
The extraplanar region defined here almost certainly contains stars that belong to
stucturally distinct components, such as a thick disk, a classical halo, tidal features, 
and possibly stellar ensembles that are associated with the nuclear outflow. 
This being said, since the WIRCam observations are 
restricted to the brightest evolved stars in NGC 253 then the analysis is biased against 
the detection of structural components that are dominated by old objects, such as the 
classical halo, and there is a bias towards structures that are populated by 
objects with ages $\leq 1$ Gyr. Indeed, the $K$ LFs of AGB stars in the 
extraplanar regions of Fields 1 and 2 are very similar (\S 5). This 
result, coupled with large scale distribution of these objects, 
which is also explored in \S 5, indicates that -- at least from a stellar population 
perspective -- the WIRCam data probe what appears to be a single extraplanar 
component. Deeper photometric studies will allow for a more complete identification 
and characterization of the other components that make up the disk and extraplanar regions 
of NGC 253.

	The main body of NGC 253 in moderately deep exposures (e.g. the DSS blue image 
shown in Figure 1) has a width of $\sim 120$ arcsec (2.3 kpc at the assumed distance of 
NGC 253), and this is consistent with the distribution of RSGs and M giants 
in Figure 3. For the purpose of investigating stellar content trends in 
the disk plane, stars that fall within a 120 arcsec 
wide strip along the major axis are assumed to be in the disk, 
and the photometric properties of these objects are discussed in \S 4.

	Stars that fall outside of the disk strip, including all of those in Field 3 and 
in the diffuse disk/extraplanar boundary zone, are assumed to be extraplanar, and 
the photometric properties of these objects are discussed in \S 5. The term `extraplanar' 
was intentionally selected to describe these objects because it is very general, and lacks 
conotations as to the origins of these objects. Referring to these 
as `halo' stars might imply to some readers that they are old and metal-poor, 
which is not the case (\S 5).

	There is cross-contamination between the disk and extraplanar samples. 
The portions of the extraplanar region that are closest to the disk strip undoubtedly 
contain some stars that belong to the classical disk, while the disk strip is 
contaminated by extraplanar stars; such contamination of the disk sample will be most 
significant for regions of the disk strip that are at large distances from the galaxy 
center, and it is demonstrated in \S 4 that this contamination is modest. 

\section{THE DISK STARS}

\subsection{The CMDs of Stars in the Disk Plane}

	The $(K, J-K)$ CMDs of sources in the disk plane of Fields 1 and 2 are shown in 
Figures 4 and 5. Each CMD contains objects in a 4 kpc wide annulus in the plane of the disk, 
with the inclination of the disk to the line of sight assumed to be 76.1 degrees. 
The distance listed in each panel, R$_{GC}$, refers to the midpoint of the annulus, as 
measured along the semi-major axis. The panels labelled `Sum' show the composite CMD 
of sources in all the distance intervals in that row, and these are shown to 
illustrate the differences between the objects that are present at R$_{GC} \leq 18$ 
kpc and at R$_{GC} \geq 22$ kpc. 

	There is substantial contamination at intermediate and larger R$_{GC}$ 
from objects that do not belong to NGC 253. Indeed, the 
number of sources arcmin$^{-2}$ is constant at large R$_{GC}$, 
as expected if the number counts are dominated by foreground stars 
and background galaxies, which should have an approximately 
uniform distribution when averaged over moderately large angular scales. 
Foreground stars populate the sequence with $K \leq 21$ near $J-K \sim 0.5$, while 
background galaxies, which are the dominant source of contamination at the magnitudes 
covered by NGC 253 stars, populate the diffuse, red plume of objects with $J-K \sim 1.2$.

	Comparisons with isochrones (\S 4.2) indicate that the majority 
of sources in the R$_{GC} \leq 18$ kpc CMDs are oxygen-rich AGB stars, and 
these form a sequence with $J-K \sim 1.2$ that peaks near $K = 19$ (M$_K 
\sim -9$). Moderately bright objects with $J-K > 1.6$ are seen to 
the right of the oxygen-rich AGB sequence. When compared with 
source counts at larger R$_{GC}$, overdensities of objects with $J-K > 1.6$
are clearly evident redward of the M giant AGB sequence with $K$ between 19 and 20 
in the 10 kpc (Field 1) and 14 kpc (Field 2) CMDs, and these red objects 
are C stars.

	The vast majority of objects above the AGB-tip in the R$_{GC} \leq 10$ kpc CMDs are 
RSGs. These stars formed during the past few tens of millions of years and are the brightest 
signatures of recent star formation in the near-infrared. There are almost no RSGs in the 
14 kpc CMDs, indicating that recent star formation in the NGC 253 
disk has largely been restricted to R$_{GC} \leq 10$ kpc.

	There are differences between the Field 1 and 2 CMDs. The most conspicuous 
difference occurs at the bright end of the CMDs, where there is a well-defined RSG sequence 
in the R$_{GC} = 10$ kpc Field 2 CMD, but many fewer RSGs, defining a less 
coherent sequence, in the R$_{GC} = 10$ kpc Field 1 CMD. The number of RSGs in the 
R$_{GC} = 6$ kpc CMDs of Fields 1 and 2 are similar, suggesting that at smaller 
radii the RSG population is more uniformly distributed. The red stellar 
content resolved with WIRCam thus indicates that stars that formed 
during the past $\sim 0.1$ Gyr are not uniformly 
distributed throughout the outer disk of NGC 253, and that there are large areas 
where the recent SFR has evidently been much higher than elsewhere in the disk.

	A more subtle difference between Field 1 and 2 is evident 
between $K = 20$ and $K = 21$ in the intermediate radii CMDs. There is a sudden drop in the 
prominence of the AGB sequence in this magnitude range between 14 and 18 kpc in 
Field 1. In contrast, the radial decrease in AGB stars in the Field 
2 CMDs is more gradual, with AGB stars apparent out to R$_{GC} = 22$ kpc in Field 2. 
These data thus suggest that the AGB content of the NGC 253 disk is azimuthally 
asymmetric, and this is investigated further in \S 4.3.

\subsection{Comparison with Isochrones}

	The only published spectroscopic study of HII regions in NGC 253 was conducted by 
Webster \& Smith (1983), and their measurements indicate that [O/H] in NGC 253 is 0.4 dex 
higher than in the LMC. Thus, comparisons with isochrones in the present work are confined 
to solar metallicities, and the R$_{GC} = 10$ kpc CMDs of Fields 1 and 2 are compared with 
solar metallicity isochrones from Girardi et al. (2002) in Figure 6. The error bars 
in the right hand panel show the $1\sigma$ scatter in the photometry predicted 
from the artificial star experiments. A distance modulus of 28.0, as measured by 
Karachentsev et al. (2003) from the RGB-tip, has been assumed, along with 
a line of sight foreground extinction A$_B = 0.05$ (Burstein \& Heiles 1982). This 
extinction is lower than that measured by Schlegel et al. (1998), which 
may be skewed by cool dust in NGC 253 (e.g. Davidge 2008c).

	The models from which the isochrones were constructed 
assume oxygen-rich atmospheres, and so follow the evolution 
of M giants, but not C stars. The isochrones indicate that the M giants with $J-K$ between 1 
and 1.5 and M$_K \geq -7.5$ and $\leq -8.5$ (i.e. $K$ between 20.5 and 19.5) have ages 
$\leq$ a few Gyr. As for the oldest stars in NGC 253, the WIRCam data sample the 
AGB component and - at least in in low density areas - stars that 
are evolving at the upper end of the red giant branch (RGB). 
The peak brightness of the AGB concentration is consistent with an age $\sim 0.1$ 
Gyr, indicating that the disk of NGC 253 has been actively forming stars throughout the 
recent past. 

	The 10 Myr and 30 Myr isochrones define a near-vertical RSG sequence near the bright 
end of the CMDs. There is a rich RSG sequence in the Field 2 CMD, 
and the range of M$_K$ occupied by these RSGs suggests that they formed $\geq 10$ Myr in 
the past. RSGs do not form in systems with ages that are much less than 10 Myr, and so 
it is worth noting that the peak of the RSG sequence in Field 2 matches that of the 10 
Myr isochrones to within a few tenths of a magnitude.

	The RSG locus in Field 2 is offset by 0.2 magnitudes in $J-K$ from the 10 Myr 
isochrone. Such a difference in color could indicate that the RSGs have a 
metallicity that is higher than solar. However, this is not consistent 
with the abundances measured by Webster \& Smith (1983) from HII regions. 
The red colors of the RSGs could also indicate uncertainties in the model colors, although 
M supergiants in the Galaxy and the Magellanic Clouds have $J-K \sim 1$ (e.g. 
Elias, Frogel, \& Humphreys 1985), which is consistent with the isochrone predictions.

	A more likely explanation for the difference in color between the models and RSGs is 
that there is substantial reddening within NGC 253. The reader may recall that only 
foreground extinction was removed from the CMDs in Figure 6, and there is evidence for 
considerable amounts of dust in the NGC 253 disk (e.g. Sofue et al. 1994). In fact, 
internal extinction provides a numerically plausible explanation of the 
color offset between the RSGs and isochrones in Figure 6. 
Pierce \& Tully (1992) apply the method described by Tully \& Fouque (1985)
to compute an internal extinction of A$_B = 0.65$ magnitudes for NGC 253. This 
statistically-based extinction estimate corresponds to $E(J-K) \sim 
0.1$ magnitude, and thus accounts for one half of the color difference seen in Figure 6. 
This is almost certainly a lower limit to the actual internal extinction of the brightest 
RSGs as young star-forming regions typically have higher levels of extinction than the main 
body of the disk, and this could amount to many tenths of a magnitude in A$_V$ (e.g. 
Zaritsky 1999). Therefore, internal extinction is a credible explanation for much (and 
possibly all) of the difference in color between the isochrones and the observed 
RSG sequence.

	The areas of the CMD that contain the samples of M giants, C stars, 
and RSGs that were used to construct Figure 3 are indicated 
in the right hand panel of Figure 6. The RSG area follows the portion of the Field 2 
R$_{GC} = 10$ kpc CMD that contains these objects, with the lower magnitude boundary 
corresponding to the approximate peak brightness of M giant AGB stars in NGC 253. 
The lower magnitude boundary for M giants at M$_K = -7$ is a 
compromise between the growing number of these objects 
towards fainter M$_K$ and the need to reduce contamination from background galaxies. This 
magnitude also corresponds approximately to the onset of thermally pulsing (TP) AGB 
evolution, and so most of the M giants plotted in Figure 3 are evolving on the TP-AGB. 
The red boundary for M giants is set at $J-K = 1.6$, which is the approximate color 
in the LMC $(K, J-K)$ CMD at which a distinct C star plume becomes evident 
(Nikolaev \& Weinberg 2002), while the blue boundary of the M giant region at $J-K = 0.8$ 
follows the blue envelope of the AGB plume. 

	The upper and lower magnitude boundaries of the C star region 
in Figure 6 bracket the magnitude range of cool C stars in the LMC, 
as defined by the C star plume in the $(K, J-K)$ CMD of that galaxy 
(e.g. Nikolaev \& Weinberg 2002). The blue limit at $J-K = 1.6$ separates M 
giants and C stars in the LMC, while a red color limit of $J-K = 2.5$ was also imposed.
The C star area in Figure 6 thus excludes `warm' C stars, 
which have $J-K$ colors that overlap with those of M giants. Lacking information 
on the depths of C-based molecules in the spectra of individual stars, 
the identification of warm C stars is problematic. However, warm C stars have intrinsically 
fainter M$_K$ than their cool brethren. Indeed, the $K-$band brightness 
of C stars in the LMC (Figure 2 of Battinelli et al. 2007) drops from 
M$_K = -7$ at $J-K = 1.5$ to M$_K = -5$ at $J-K = 1$, which places these stars below the 
faint limit of the WIRCam observations. The omission 
of warm C stars thus does not have a major impact on the luminous end of the C star LF.

\subsection{The Spatial Distribution of Stars in the Disk Plane}

\subsubsection{M Giants}

	The LFs of stars with $J-K$ between 0.8 and 1.6 in the disk planes of Fields 
1 and 2 are compared in Figure 7. While this color range was selected 
primarily to accomodate M giants, RSGs are also included at the 
bright end of the LFs. The LFs in Figure 7 were corrected for foreground stars 
and background galaxies by subtracting source counts from the R$_{GC} = 30$ and 34 kpc 
annuli. 

	The relation between the peak M$_K$ of the AGB and age predicted from 
the solar metallicity models of Girardi et al. (2002) is shown 
at the top of Figure 7. The Field 1 and 2 M giant LFs at R$_{GC} = 6$ and 
10 kpc have comparable numbers of stars at M$_K = -8.5$ and --8. 
Given that the age calibration suggests that AGB stars with 
M$_K = -8.5$ formed $\leq 2$ Gyr in the past, then it 
appears that stars that formed during intermediate 
epochs are uniformly distributed throughout the inner regions of the NGC 253 disk. 
However, there are obvious differences between the Field 1 and 2 R$_{GC} = 6$ and 
10kpc LFs at the bright end, where RSGs dominate the number counts. The number of RSGs at 
R$_{GC} = 10$ kpc in Field 1 is almost an order of magnitude lower than at the same R$_{GC}$ 
in Field 2, confirming the impression obtained from the CMDs and Figure 3 that the youngest 
stars in the disk of NGC 253 are not distributed symmetrically throughout the disk. 

	Evidence for asymmetries in the NGC 253 disk is also found at larger R$_{GC}$. 
The number densities of M giants at R$_{GC} = 14$ and 18 kpc 
in Field 2 exceeds those in Field 1 by 0.4 -- 0.6 dex. 
The offset between the Field 1 and 2 LFs is roughly constant over the 
range of M$_K$ probed with these data, indicating that the difference in number counts 
is not due solely to the youngest stars. This indicates that while the AGB 
population in the {\it inner} disk of NGC 253 is 
distributed symmetrically, the number density of AGB stars in the {\it outer} disk is 
lop-sided, in the sense of significantly higher densities in Field 2. 

	With the exception of the bright end of the Field 2 LF, the shape of the disk plane 
LFs do not change with radius. Neglecting the systematic differences in the number density 
of stars at R$_{GC} = 14$ and 18 kpc, then within the (large) uncertainties at the bright 
end of the LFs with R$_{GC} \geq 14$ kpc the overall shapes of the Field 1 and 2 LFs 
are similar. The calibration at the top of Figure 7 suggests that the brightest 
M giants in the R$_{GC} = 14$ and 18 kpc annuli of both fields have ages $\leq 0.5$ Gyr. 

	The R$_{GC} = 22$ and 26 kpc LFs were 
combined to boost the S/N ratio in these regions, and significant numbers 
of stars are seen in the resulting composite LF. The majority of stars detected in the 
R$_{GC} = 22$ -- 26 kpc interval have M$_K \geq -7$, and 
this is consistent with them being upper RGB stars in moderately metal-poor ([M/H] $> -1$) 
old populations, although they may also be intermediate age AGB stars. The
agreement between the number counts in the Field 1 and Field 2 $22 + 26$ kpc LFs 
indicates that the lop-sided mass distribution may not extend to the outermost regions 
of the disk.

	Because they are in a short-lived stage of evolution, and as such are relatively 
rare, the detection of the most evolved AGB stars becomes progressively more problematic 
with increasing R$_{GC}$. This introduces a radial bias in the age estimates, in the 
sense of overestimating the age of the youngest population as one goes to larger R$_{GC}$. 
If the R$_{GC} = 14$ kpc LFs are scaled to match the numbers of stars with M$_K 
\sim -7$ at R$_{GC} = 22$ and 26 kpc, then it is evident that bright AGB stars 
would not be expected in statistically significant numbers. Therefore, 
the possibility that the disk of NGC 253 at very large 
R$_{GC}$ may contain stars that formed within the past few 100 Myr 
can not be excluded with the WIRCam data. 

	The distribution of bright M giants in the disk plane is examined 
further in Figure 8, where n$_{TAGB}$ is the number of stars per arcmin$^{2}$ 
with $K$ between 19.5 and 20.5 (i.e. M$_K$ between --8.5 and --7.5) and $J-K$ between 
0.8 and 1.6. The difference in the number density of M giants in Field 1 and 2, which was 
noted previously, is clearly evident, and the number counts in Field 2 
exceed those in Field 1 by $\sim 0.5$ dex at all R$_{GC}$. 
In addition, the Field 2 number counts define an exponential 
profile that continues to R$_{GC} = 23$ kpc ($\sim 14$ disk scale lengths). This is in 
marked contrast to the Field 1 number counts, 
which follow only an ill-defined exponential, with 
a plateau in the number counts at R$_{GC} = 15 - 17$ kpc.

	In \S 5 it is demonstrated that there is a diffuse extraplanar 
component, and some of the stars at large R$_{GC}$ in the disk sample may be
stars that are not in the disk plane. However, the majority of stars detected at this 
large R$_{GC}$ in the disk sample likely are not extraplanar. Indeed, 
source counts discussed in \S 5 indicate that the number density of extraplanar 
stars in Field 2 is $\sim 0.7$ dex lower than in the Field 2 disk plane at roughly the 
same distance from the galaxy center, and an even greater difference in number density 
occurs when comparing with the disk and extraplanar LFs of Field 1. Thus, 
the highest stellar densities at large distances from the center of NGC 253 are found 
in the disk plane. In summary, the WIRCam data indicate that the disk of 
NGC 253 extends out to distances of at least 22 -- 26 kpc from the galaxy center. 

\subsubsection{C Stars}

	C stars provide additional insights into the star-forming 
history of NGC 253. While sharing a common physical characteristic -- an atmosphere 
in which [C/O] $\geq 0$, which results in an absorption spectrum that is dominated by 
C-based molecules -- C stars are a heterogeneous group of objects from a stellar content 
perspective, as they can have ages that range from 0.2 to 3.2 Gyr. A further complication 
is that the fraction of the time during TP-AGB evolution 
that a star spends as a C star depends on metallicity, since 
as one moves to higher metallicities then more C must be dredged up to bind atmospheric 
O. The fraction of the TP-AGB phase that is spent as a C star thus 
drops as metallicity increases. The statistics of C stars in the LMC suggest that stars 
with Z = 0.008 that formed 0.45 -- 2.5 Gyr in the past will spend at least 50\% of their 
TP-AGB evolution as a C star. For comparison, stars with solar metallicity in the same age 
range will typically spend only $\sim 30\%$ of their TP-AGB evolution as C stars (e.g. 
Maraston 2005).

	Bolometric magnitudes for C stars were computed using 
$K-$band bolometric corrections from Bessell \& Wood (1984) for Galactic and LMC 
stars, and the bolometric LFs of C stars in Fields 1 and 2 are compared in Figure 
9. The relation between peak AGB M$_{bol}$ and age that is 
predicted by the solar metallicity models of Girardi et al. (2002) is shown at the top of 
Figure 9. While the models assume O-rich atmospheres, the intrinsic luminosities of 
O-rich and C-rich stars with the same age and mass should be comparable. 
Due to small number statistics, the brightest C stars give only a lower limit to the 
age of a system, and stars with M$_{bol}$ that correspond to that of the AGB peak for 
a given age in the Girardi et al. (2002) calibration must be that age or younger. 

	The C star LFs of disk stars in Fields 1 and 2 at R$_{GC} = 6$ are in excellent 
agreement, echoing the agreement between the M giant LFs at M$_K \sim -8$ noted previously. 
Such a uniform distribution among stars with ages 1 -- 3 Gyr is perhaps not surprising, 
as stars with this age have been in place for many disk rotations, and thus might be 
expected to be well-mixed throughout the NGC 253 disk.
This being said, the shape of the disk C star LF changes with R$_{GC}$ in both 
fields. The R$_{GC} \leq 10$ kpc C star LFs are relatively flat or climb to higher 
numbers at the bright end. However, when compared with the LFs at smaller R$_{GC}$, 
the R$_{GC} = 14$ kpc LFs are deficient in objects at the bright 
end, while the Field 2 LF at R$_{GC} = 18$ kpc has a distinct downward tilt towards 
brighter M$_{bol}$. This change in LF characteristics likely signals an age gradient in the 
NGC 253 disk, in the sense that the mean age of C stars increases towards larger R$_{GC}$.
With the caveat that the measurement of the AGB-tip is problematic when only small 
stellar samples are available, the brightest M giants in the Field 1 and 2 $22 + 26$ kpc 
LFs have ages $\geq 1 - 2$ Gyr if they are evolving near the AGB-tip.

	There is a systematically higher density of C stars in Field 2 than in Field 1 
at R$_{GC} > 6$ kpc, and this difference grows with increasing R$_{GC}$. At R$_{GC} = 14$ 
kpc the number density of C stars is 0.5 dex higher in Field 2 than in Field 1, which is 
in line with the difference between the M giant LFs of these fields at the same R$_{GC}$. 
At R$_{GC} > 14$ kpc C stars have all but disappeared in Field 1, whereas significant 
numbers are seen in Field 2. M giants are seen in significant numbers out to R$_{GC} = 22 
- 26$ kpc in Field 1, with number densities that are comparable to those in Field 2 at 
large R$_{GC}$ (see Figure 7). If the stellar content in Field 1 is like that in Field 
2 then C stars should be seen at large radii in Field 1, but none are detected.

	Why do C stars disappear in Field 1 at large R$_{GC}$? Two general explanations 
for the change in C star counts are considered here: star-forming history and metallicity.
The discussion starts with the assumption that the C stars formed {\it in situ} in the 
NGC 253 disk, although the prospect of them forming in a companion is discussed 
at the end of the section.

	Assuming a constant SFR and a moderately high metallicity for NGC 253, then 
the majority of C stars formed $1 - 2$ Gyr in the past (Figure 12 of 
Maraston 2005). A dearth of C stars at R$_{GC} > 14$ kpc in Field 1 could then result 
if the eastern half of NGC 253 is deficient in stars that formed 1 -- 2 Gyr in the past 
when compared with the western half of the disk. If this was the case then M giants 
in Field 1 at R$_{GC} = 18$ kpc would have formed predominantly within the past 1 Gyr, 
and the M giant locus in the outer regions of Field 1 would then have a bluer 
ridgeline color than in Field 2. In fact, the M giant locus in Field 1 
{\it is} matched approximately by the 0.1 Gyr isochrone 
in Figure 6, while in Field 2 the M giant locus roughly coincides with the 1 Gyr isochrone. 
Alternatively, such a difference in color could also be due to systematically 
higher reddening in Field 2 when compared with Field 1.

	Metallicity is another quantity that plays a key role in defining C star statistics. 
However, metallicity is a cumulative quantity that is the result of Gyr of evolution. 
To explain the difference in C star numbers noted above, the 
star-forming history of Field 1 would have had to have differed from that of Field 2 
for a long period of time if the disk of NGC 253 has evolved as a closed system. This 
seems unlikely given that mixing will occur throughout the disk during Gyr timescales. 

	It is not known if the C stars in the outer regions of NGC 253 formed 
in NGC 253, or in an object that was subsequently accreted by NGC 253. If the latter 
is correct, then the bulk of the satellite was presumably accreted into Field 2 
and has not yet been mixed throughout the galaxy (i.e. the object was 
accreted within the last Gyr). The difference in C star content between Fields 1 and 2 
could then be due to either star-forming history and/or metallicity. Metallicity becomes 
a much more viable explanation for the dominant cause of 
C star differences if the stars at large R$_{GC}$ in Field 
2 were either accreted from a companion galaxy or formed from gas that was 
accreted from another galaxy, as the chemical evolution of that system was almost 
certainly very different from that of the NGC 253 disk.

\section{THE EXTRAPLANAR STARS}

\subsection{The CMDs}

	The $(K, J-K)$ CMDs of sources in Field 3 are shown in Figure 10, 
while the CMDs of the extraplanar objects in Fields 1 and 2 are shown in Figures 
11 and 12. As in Figures 4 and 5, the CMDs include sources in 4 kpc wide intervals. 
However, unlike Figures 4 and 5, the midpoint radius of each annulus shown in Figures 
10 -- 12 is the radial distance from the center of the galaxy measured on the plane 
of the sky, $z_{dp}$, as opposed to distance measured in the plane of the disk. 
Composite CMDs of all sources in the CMDs in each row are also shown. 

	The general appearance of the CMDs in Figures 10 -- 12 changes gradually with 
$z_{dp}$, transitioning from CMDs that are dominated by NGC 253 stars to those that 
are dominated by foreground stars and background galaxies. The CMDs with $z_{dp} \leq 
10$ kpc have a prominent plume at $J-K \sim 1.2$ and with $K 
\geq 20$ that is due to M giants, and a modest concentrations of M 
giants is seen in the CMDs out to $z_{dp} \sim 14$ kpc. 
There is also a spray of objects with $J-K > 1.6$, that is most obvious in the 
$z_{dp}  = 10$ kpc CMD, and many of these are cool C stars. 
The CMDs in Figures 10 -- 12 thus indicate that an intermediate age extraplanar 
component can be traced out to large $z_{dp}$. In addition, that the extraplanar component 
extends from Field 3 into Fields 1 and 2 indicates that intermediate age extraplanar 
stars are not restricted to the minor axis of NGC 253, but likely surround the galaxy.

	While all sources in Field 3 have been classified as extraplanar, 
there is some ambiguity in the physical location within NGC 253 of stars in Field 3. 
The orientation of NGC 253 on the sky is such that disk stars huddle along a 
relatively narrow strip, and the amount of disk contamination 
in the extraplanar regions drops rapidly with increasing z$_{dp}$, as 
the density of disk stars drops and progressively larger extraplanar volumes are sampled. 
Still, some of the bright AGB stars in the fields 
with the smallest $z_{dp}$ in Field 3 probably belong to the 
disk. If the classical disk of NGC 253 extends out to R$_{GC}  = 22$ kpc, as 
suggested by star counts in Fields 1 and 2, then some objects that have $z_{dp} \sim 5$ kpc 
near the minor axis in Field 3 may belong to the classical disk, as opposed to 
being extraplanar, but these will be modest in number given the density of objects 
in this part of teh disk (Figure 8).

\subsection{Comparisons with Isochrones}

	The CMDs of sources in Field 3 with very different $z_{dp}$ are compared 
with Z = 0.019 isochrones from Girardi et al. (2002) in Figure 13. As with the comparisons 
in Figure 6, a correction was applied only for foreground extinction, and not internal 
extinction. Given the orientation of NGC 253 on the sky, a small number of sources in the 
$6 + 10$ kpc CMD may belong to the classical disk. The CMD of sources with z$_{dp} = 
30 + 34$ kpc is also shown to allow the reader to assess by eye the contamination from 
foreground stars and background galaxies in the $6 + 10$ kpc CMD. The $6 + 10$ kpc 
and $30 + 34$ kpc intervals sample similar total areas on the sky.

	When compared with the $30 + 34$ kpc CMD, the $6 + 10$ kpc CMD 
contains a clear excess population of objects with M$_K > -8.5$ and $J-K$ between 1 and 1.5. 
The 1 Gyr isochrone passes through the middle of this population of objects, 
which are M giants in NGC 253. The peak of the M giant sequence in the $6 + 10$ kpc interval 
indicates that -- as in the disk of NGC 253 -- the extraplanar regions of NGC 253 
contain large numbers of stars that formed at least 1 Gyr in the past. A well-defined RSG 
sequence is not seen in the $6 + 10$ kpc CMD, and so there are no signs of 
very recent star formation in the WIRCam data at these $z_{dp}$.

\subsection{The Properties of Luminous Extraplanar AGB Stars}

\subsubsection{M Giants}

	The $K$ luminosity functions (LFs) of extraplanar objects with 
$J-K$ between 0.8 and 1.6, which is the color range containing bright M 
giants, are shown in Figure 14. The Field 1, 2, and 3 LFs are plotted with solid, 
dashed, and dotted lines, respectively. As in \S 4, the LFs have been corrected 
statistically for contamination from Galactic foreground stars and 
background galaxies by subtracting source counts at large offsets from the galaxy center. 

	That sources are seen in Field 3 in significant numbers out to $z_{dp} = 14$ kpc 
indicates that an extraplanar component is present, and extends to large distances off 
of the disk plane, in agreement with the visual impression obtained from 
the CMDs. In addition, there is considerable field-to-field variation in the 
number density of extraplanar stars at the same z$_{dp}$. The spatial distribution 
of extraplanar stars is skewed to higher values near the disk plane, with the 
highest concentration of objects in Field 2. Such field-to-field
differences are suggestive of either a lumpy stellar distribution or a 
departure from the circular symmetry that is assumed to compute $z_{dp}$. 

	The number densities of stars with M$_K \geq -8$ in Field 2 tend to be a few 
tenths of a dex higher than in Field 1 at all R$_{GC}$, while the number density of 
sources in Field 3 is roughly an order of magnitude lower than in Fields 1 and 2. 
The agreement between the source densities in Fields 1 and 2 suggests that localized 
density variations may account for differences in number counts of no more than a 
few tenths of a dex when averaged over kpc scales. As for a non-cicular projected 
distribution, the tendency for higher number densities to occur in Fields 1 and 2 
at a given $z_{dp}$ is indicative of an elliptical distribution, in broad agreement 
with the overall visual distribution of stars in Figure 3. That the number density of stars 
with $K$ between 20 and 21 at $z_{dp} = 14$ kpc in Field 3 matches the number density of 
stars in the same magnitude range at $z_{dp} = 22 - 26$ kpc in  Fields 1 and 2 suggests that 
the projected ellipticity of the extraplanar component is $1 - \frac{14}{24} \approx 0.4$.

	The difference in number densities of extraplanar sources between 
Fields 1 and 2 mirrors the distribution of disk 
stars, hinting that the disk plane and extraplanar populations identified in this paper 
are not completely de-coupled, and might have a common heritage. 
The degree of similarity between the LFs of stars in the disk plane and the 
extraplapanar regions is investigated in Figure 15, where the extraplanar LFs of Fields 
1 and 2 are compared with the R$_{GC} = 14$ kpc disk plane LFs of M giants in the same 
fields. The disk plane LFs have been scaled to match the mean number counts 
in the extraplanar LFs at M$_K = -7.5, -8.0$, and --8.5. The 
higher stellar density in the disk elevates the completeness limit 
in the disk plane LFs with respect to that in the extraplanar regions (\S 2). 

	Field-to-field differences in completeness 
notwithstanding, the disk plane and extraplanar LFs have similar shapes at 
the bright end, signalling similar star-forming histories over a large fraction 
of the age of the Universe. The agreement between the disk and extraplanar LFs in 
Figure 15 is consistent with a disk origin for the extraplanar stars. If this is correct 
then the disk must have been disrupted not too far in the past, given the 
similarities between the disk and extraplanar LFs at the bright end.

	The spatial distribution of extraplanar M giants in NGC 253 provides further 
insights into their nature and origins. The number density of AGB stars with 
$K$ between 19.5 and 20.5 (i.e. M$_K$ between --8.5 and --7.5) and $J-K$ between 0.8 
and 1.6 is shown in Figure 16 as functions of z$_{dp}$ (top panel) and z$_{dp}^{1/4}$ 
(lower panel). A linear relation in the top panel is indicative of an exponential 
distribution, as is associated with rotationally-supported disks, while a linear relation in 
the lower panel is indicative of a distribution that is associated with a classical 
non-rotating virialized system.

	There are only subtle differences between the distribution of points in the two 
panels of Figure 16. While a linear relation with a small amount of scatter can be fit to 
the points in the top panel, a reasonable linear fit can also be made to the points in the 
lower panel, albeit with larger scatter than in the top panel. These data 
thus lack the statistical power to characterize unambiguously 
the spatial distribution of the brightest M giants in this 
part of NGC 253. The structural characteristics of the 
brightest M giants notwithstanding, the absence of marked discontinuities in the 
distribution of the brightest M giants suggests that these stars belong to 
a single population of objects that is well-mixed throughout the extraplanar regions of 
NGC 253.
 
\subsubsection{C Stars}

	A population of objects with $J-K > 1.6$ 
is seen throughout the extraplanar regions, and some fraction of these 
are luminous cool C stars. The M$_{bol}$ LFs of these objects, 
constructed following the procedures described in \S 4, including a statistical correction 
for contamination from foreground and background sources, are shown in Figure 17. 
As with the M giant LFs, Field 2 tends to have the highest density of objects. 
With the caveat that the uncertainties in the Field 1 
and Field 3 LFs tend to be substantial at most $z_{dp}$, 
the density of objects in Field 3 tends to be comparable to 
that in Field 1. The C stars thus appear to have a more 
circular distribution on the sky about the center of NGC 253 than the luminous M giants; 
a survey for C stars along the north west minor axis will provide additional information 
to test the veracity of this result. The C star LFs of all three fields have similar shapes, 
that appear not to change with $z_{dp}$. The age calibration at the top of 
Figure 17 suggests that the youngest C stars throughout much of 
the extraplanar regions have ages $\leq 1$ Gyr, further 
supporting the notion that the regions surrounding NGC 253 contain stars that 
formed during cosmologically-recent epochs.

\subsection{The Star-Forming History of the Extraplanar Regions}

	The relation between peak AGB brightness and age that is shown at the top of Figure 
14 indicates that subtantial numbers of stars with ages $< 1 - 2$ Gyr are present throughout 
the extragalactic regions out to projected distances from the galaxy center that are 
comparable to that of the physical extent of the disk along the major axis. 
The ages of the youngest stars, which could be gauged from the peak brightness of the 
AGB, is of obvious interest, but this is a difficult measurement to 
make. Small number statistics are an issue at $K \leq 19.5$ at 
almost all $z_{dp}$, although there are significant numbers of stars with 
magnitudes that are consistent with an AGB-tip age of 0.5 Gyr at small $z_{dp}$. 

	While providing information about the age of the youngest AGB stars, 
the AGB-tip has limitations as a probe of stellar content. 
The use of the AGB-tip as the sole age diagnostic is inefficient, 
in the sense that other age-related information is ignored.
In addition, determining the magnitude of the AGB-tip can be affected by small 
number statistics, photometric variability, and the potential for obscuration by thick 
circumstellar disks, which introduces a potential bias against the detection of the 
intrinsically most luminous stars in surveys at visible and near-infrared wavelengths. 

	While not providing a pancea for overcoming the shortcomings associated with 
the AGB-tip, the LF of AGB stars offers additional information about 
the mix of stellar ages within a system. The LF also has the merit of using all 
stars in a sample that are brighter than some magnitude cut-off, as opposed to only those 
that are the very brightest. In this study, insights into the star-forming history 
of the extraplanar regions of NGC 253 are obtained by comparing the observed LFs 
with model LFs that were constructed from routines in Starfish (Harris \& Zaritsky 2001).
The goal of this effort is not to model specific features in the observed LFs, but rather 
to generate LFs that will serve as interpretive guides. 

	The simplest star-forming history is a single episode of star formation that forms 
stars having the same metallicity - a so-called simple stellar population (SSP). If the 
bright extraplanar stars in NGC 253 formed {\it in situ} as the result of a single event 
then they might be expected to have only a modest dispersion in age and metallicity, and hence 
the LF of these stars could be approximated by that of a SSP. Model LFs for SSPs generated 
from the solar metallicity Girardi et al. (2002) isochrones are compared in Figure 18 
with the mean LF of sources in the $z_{dp} = 10$ and 14 kpc intervals of Field 3. 
A Salpeter mass function has been assumed, although the model LFs in the magnitude range 
considered here are not sensitive to the properties of the mass function. 

	The Field 3 LF is steeper than the SSP model LFs. This 
departure from flatness is most noticeable when M$_K \leq -8$, and 
indicates that the AGB stars in the extraplanar regions of NGC 253 have a range of ages. 
Thus, the extraplanar stars are not part of an SSP that formed within, say, the past 
$\sim 1$ Gyr.

	Another interesting reference model is that of a single-metallicity system 
that has experienced a continuous star-forming history with a constant SFR. This scenario is 
highly idealized and does not account for chemical evolution. Still, the star-forming 
activity of spiral galaxy disks and dwarf irregular galaxies with M$_B < -15$, which 
is the luminosity range in which stochastic flucuations in star-forming 
history are greatly reduced (e.g. Weisz et al. 2008), is more-or-less continuous. 
The model labelled `Constant SFR' in Figure 18 shows the LF of a solar metallicity system that 
has experienced a constant SFR over the age range 0.2 -- 10 Gyr. 
This model predicts decreasing star numbers towards more luminous M$_K$, 
and the slope of the model LF is roughly consistent with that of the $10 + 14$ kpc Field 3 LF.

	The agreement between the observations and the Constant SFR 
model LF is not perfect, and it should be recalled that no effort has been made 
to finesse the star-forming history to match the observations. For example, 
the break in the $10 + 14$ kpc LF near M$_K = -8.25$ could be the result 
of a drop in the SFR $\sim 1 - 2$ Gyr in the past. Other astrophysically relevant improvements 
would be to include (1) a large departure from a constant SFR at very 
early epochs to account for the elevated SFR that is associated with the initial stages of 
galaxy assembly, and (2) the effects of chemical enrichment. However, neither of these 
additions to the models would change the basic conclusion 
that can be reached from the comparisons in Figure 
18: the extraplanar regions of NGC 253 contain stars that span a range of ages. Including 
a SFR spike at early epochs would only impact the two faintest bins plotted in Figure 18. 
As for the inclusion of chemical enrichment, the LF is dominated by stars with ages 
$\leq 1$ Gyr. Therefore, only very steep age-metallicity relations would have a significant 
impact on the models, and a steep age-metallicity relation is 
not consistent with observations of nearby spiral galaxies (e.g. Magrini et al. 2009). 
The potential for fine-tuning the model LFs aside, it is clear that the LF of extraplanar 
stars is consistent with them not being part of a coeval system; rather, the extraplanar 
regions of NGC 253 are populated by stars that formed over a protracted period of time.

\section{DISCUSSION \& SUMMARY}

	Deep $J$ and $K$ images obtained with the CFHT WIRCam have been used to explore 
the stellar content in the outer regions of the nearby starburst galaxy NGC 253. This is the 
first panoramic stellar survey of this galaxy. Photometrically-selected samples 
of RSGs, M giants and C stars are used to trace the spatial distribution of stars 
with young and intermediate ages. Stars are assigned to the disk or the extraplanar regions 
based on their projected location on the sky (1) to assess population characteristics, 
and (2) to investigate spatial trends and search for structures. While there is inevitable 
contamination between the disk and extraplanar samples, this is only an issue at modest 
offsets from the disk plane, near the boundary separating the disk and extraplanar regions.

	The brightest disk stars in the near-infrared are RSGs, and comparisons with 
isochrones suggest that the youngest of these have ages $\sim 10$ Myr in NGC 253. 
RSGs are highly effective probes of recent star formation by virtue of their 
intrinsic brightness, and the distribution of RSGs in the NGC 253 disk is lop-sided, with 
many more in the north east quadrant of the galaxy than in the south west quadrant.
Unfortunately, small number statistics restrict the usefullness of RSGs as stellar 
content probes in low-density environments.

	M giants that are evolving on the AGB are more plentiful in composite 
stellar systems than RSGs, and hence provide a better means of tracing diffuse 
structures. The photometric depth of the WIRCam data is such that 
M giants with ages that span the lifetime of the galaxy are present in these data, 
although the vast majority of M giants detected here probably have 
ages $< 2 - 3$ Gyr. While M giants have the merit of being comparatively easy to 
detect because of their intrinsic luminosity, the ability to use them 
to identify individual star-forming episodes diminishes towards progressively older ages. This 
is because the rate of change of AGB-tip brightness with time decreases with increasing
age, making efforts to isolate individual large-scale star-forming events 
that occured more than a few Gyr in the past difficult. An added 
complication is that a significant fraction of stars near the AGB-tip are LPVs, 
and photometric variability blurs information in the LFs. Because of their comparatively 
high numbers, M giants have been adopted as the main tracer of structure in this paper.

	C stars provide insights into the star-forming history of a system that are 
complementary to those provided by M giants. Whereas the brightest M giants have ages 
$\sim 0.2$ Gyr, the majority of C stars have ages 1 -- 3 Gyr (e.g. Maraston 2005). That 
C stars are seen over the same spatial intervals as M giants throughout 
the disk and extraplanar regions of NGC 253 provides direct evidence that the 
disk and extraplanar regions of this galaxy contain stars that span a wide range of ages.

\subsection{The Disk of NGC 253, and Clues to its Past History}

\subsubsection{The Early Evolution of NGC 253}

	The extent to which NGC 253 can serve as a template for understanding more 
distant starburst systems depends on its nature prior to the events that 
sparked the current starburst -- if NGC 253 has not had a
`typical' history then it may not be a good comparison object for probing `typical' starburst 
activity. In fact, the properties of the globular cluster system hint that NGC 253 may 
not be a textbook spiral galaxy in terms of its past evolution. 
Expanding upon the sample of clusters identified by Beasley \& Sharples 
(2000), Olsen et al. (2004) used deep spectra to identify NGC 253 globular clusters 
and study their kinematics. Olsen et al. (2004) find that the globular clusters system of 
NGC 253 is remarkable in two respects. First, it appears to be rotationally 
supported, with a characteristic scale length that is comparable to that of 
the present-day HI disk, although there are admittedly large uncertainties in this quantity.
While NGC 253 is not unique in having a rotationally-supported globular cluster 
system, it is the best characterized such system (Olsen 2010, private communication). 
Second, the specific frequency of globular clusters in NGC 253 is more than an order 
of magnitude lower than in other Sculptor group spiral galaxies.

	That the characteristic size of the star-forming disk of NGC 253 has evidently 
not changed since early epochs is surprising given the evidence 
for inside-out disk growth in spiral galaxies (e.g. Trujillo \& Pohlen 2005; 
Colavitti et al. 2009). If disks form from the inside-out then globular clusters that formed 
early-on should have a more compact distribution than the present day 
stellar disk (at least assuming that the stellar disk of NGC 253 has not been 
tidally pruned, which appears not to be the case given the large physical extent of 
the stellar disk defined by M giants -- see below). A core assumption is that all of the 
clusters formed as part of NGC 253, and were not accreted by an interaction 
with a companion. If NGC 253 did accrete clusters then the companion 
must have been on a coplanar orbit in order to retain a disk-like cluster distribution in the 
final merged system. In any event, cluster accretion seems unlikely given 
the low globular cluster specific frequency of NGC 253.

	The presence of a spatially-extended disk-like globular cluster system 
is also surprising given that major mergers can have a highly disruptive 
effect on the properties of stars in disks prior to the merger. 
The vast majority of nearby spirals appear to have experienced such an event during 
the past few Gyr (Hammer et al. 2007). Hopkins et al. (2009) 
investigate the impact of major (mass ratios $> 0.1$) mergers on 
gas-rich disks, and find that while the initial disk of the more massive galaxy may be 
obliterated, it re-forms on a cosmologically short time scale. Existing stars are 
scattered out of the plane of the primary galaxy, and these 
subsequently virialize to form a pressure-supported system. An old 
globular cluster system in a disk galaxy that has experienced a major merger should then be 
pressure-supported, rather than distributed in a disk. The disk-like globular cluster 
distribution in NGC 253 is counter to this expectation, unless it experienced a major 
merger with a companion with an orbit that was coplanar with the NGC 253 disk. Such a merger 
geometry is conducive to producing starburst activity and maintaining asymmetries in the disk 
(such as those discussed in \S 4) if the companion orbit is prograde (Hopkins et al. 2009).

	The Sculptor group is dynamically unevolved (Karachentev 2005), and NGC 253 may 
simply be one of the $\sim 7\%$ of nearby spirals that have avoided a major merger 
(Hammer et al. 2007). Major mergers involving gas-rich systems produce 
environments that are conducive to globular cluster formation (e.g. Ashman 
\& Zepf 1992). Therefore, an uneventful interaction history could explain 
the low specific frequency of globular clusters in NGC 253.

	If NGC 253 has not experienced a major merger for a large fraction of the Hubble time 
then it might have a relatively compact stellar disk when compared with other 
nearby galaxies, due to a relative deficiency in angular momentum (Hammer et al. 2007). M81 
has an M$_K$ that is comparable to that of NGC 253, and the disk scale length of NGC 
253 (1.7 kpc; Forbes \& Depoy 1992) is comparable to that measured for M81 
from the $K-$band surface brightness profiles given by Jarrett et al. (2003). 
The WIRCam data also indicate that the physical extent of the NGC 253 disk is not compact. 
AGB stars in the disk are traced out to at least 22 kpc along the major axis (\S 4), which 
corresponds to 13 scale lengths. Given the diffuse appearance of the outer disk in 
Figure 3, it is possible that the stars seen at large R$_{GC}$ may not be in the 
classical disk; rather, they could be the result of tidal interactions. 
Still, that stars are detected out to this distance in both Fields 1 and 2 with comparable 
number densities argues that the majority of these stars are part of a dynamically mature 
component, which we associate with the disk. The extent of 
the stellar disk in NGC 253 is thus comparable to the distances 
out to which stars have been resolved in NGC 247 (Davidge 2006) and M81 (Davidge 2009), 
and exceeds the stellar disk of NGC 2403 (Davidge 2007).

	We conclude that the stellar disk of NGC 253 is 
not compact when compared with other nearby galaxies. If, as the other evidence suggests, 
NGC 253 avoided mergers up to the recent epochs, then NGC 253 may have 
received angular momentum from torques exerted by NGC 247 (Whiting 1999). 
In fact, the disk of NGC 253 is likely even more extended than indicated by AGB stars. 
The Holmberg radius of NGC 253 is 17.6 arcmin (Puche et al. 1991), or 21 kpc. Bright 
AGB stars are relatively rare, and deeper images that sample stars on the 
upper portions of the RGB, will probably trace the stellar disk of NGC 253 out to even larger 
radii. Blue main sequence stars may also provide a powerful means of plumbing the 
outer limits of the NGC 253 disk. The main sequence phase of evolution has a much longer 
duration than the AGB phase, so that main sequence stars will occur in much larger numbers 
than AGB stars. Moreover, the use of blue stars also results in greater contrast with 
respect to background galaxies, the majority of which have red colors (e.g. Davidge 2008d).

\subsubsection{Recent Disk Evolution: Evidence for an Interaction}

	The spatial distribution of recent star formation can be used to probe the 
recent history of the galaxy. Starbursts are most commonly associated with galaxy-galaxy 
interactions and so the disk of NGC 253 might be expected to harbour signatures of 
a tidal encounter. In fact, evidence for such an interaction is evident in the 
sub-structure close to the disk plane (Figure 3). 

	If there was an interaction then it evidently produced elevated levels of star 
formation not only in the inner regions of the galaxy, but also in the disk. Indeed, 
the high density of RSGs in the north east portion of NGC 253 indicates that 
this was an area of very active star formation within the past few tens of Myr. This 
interpretation is supported by the presence of NUV and FUV emission (Hoopes et al. 2005) and 
intense H$\alpha$ emission (Hoopes et al. 1996) in this part of the galaxy. 
The bright H$\alpha$ emission indicates that the elevated star-forming activity 
heralded by the RSGs almost certainly continues up to the current time. While the south west 
part of the disk is not devoid of H$\alpha$ emission and UV-bright objects, the sources 
there are fainter and less numerous than at the other end of the major axis. The north east 
portion of the NGC 253 thus appears to have had the highest levels of star formation in the 
main body of the disk during at least the past few tens of Myr. This part of 
the galaxy also shows an enhancement in the number of bright AGB stars with respect to 
the south east arm of the major axis (\S 4), suggesting that excess star formation in this 
area has occured for hundreds of Gyr.

	While the orbital properties of the supposed interacting galaxy 
are not known, there are clues to possible trajectories. The distribution of 
stars in Figure 3 is suggestive of an encounter involving a companion with an orbit that 
was close to the NGC 253 disk plane, and this is consistent with the flattened distribution
of extraplanar M giants measured in \S 5. Dynamically de-coupled 
material near the galaxy center (Anantharmaiah \& Goss 1996; Prada et al. 1998) suggests 
that the encounter may also have involved the central regions of the galaxy.
The range of interaction geometries investigated by Hopkins et al. 
(2009) indicate that the most pronounced asymmetries 
in the disk result from prograde encounters in which the companion orbit is 
close to that of the disk plane of the larger galaxy. The 
lop-sided nature of the NGC 253 disk thus suggests that the galaxy 
that interacted with NGC 253 was probably not on a retrograde orbit.

	We are not aware of published studies that investigate an interaction 
between NGC 253 and a possible companion. However, interactions between 
M31, which has an integrated brightness that is similar to that of NGC 253 and is 
known to be surrounded by tidal features, and a companion have been examined. 
There is some similarity between the general appearance of M31 with its tidal structures in 
Figure 1 of Fardal et al. (2007) and NGC 253 in Figure 3 of the current paper.

	Fardal et al. (2007; 2008) investigate the interaction between a single satellite and 
the M31 disk. The main emphasis of these studies was to model the tidal features that have 
been found near M31, and the orbital pericenter of the companion is a few kpc (Fardal 
et al. 2006). In addition to producing the M31 giant stellar stream, these models 
produce structures that are close to the observed disk plane. While originating in 
the satellite, the stars in these structures have a spatial distribution 
that is defined largely by the M31 gavitational potential. 
If the extraplanar population in NGC 253 is analogous to that in M31 then the 
Fardal et al. (2007; 2008) simulations indicate that the extraplanar regions of NGC 253 
are populated predominantly by stars that originated in a companion galaxy.

	An interaction with a satellite will also affect the stellar distribution 
in the M31 disk. Block et al. (2006) consider a companion (presumably M32) that passes 
through the center of M31, and that study provides insights into signatures 
in the disk that such an encounter might produce. The interaction in that model 
produces a density wave that propogates outwards from the point of impact. This wave spurs 
star formation as it moves along, and localized regions of star formation in 
ring- or arm-like structures result (Figure 2 of Block et al. 2006). 

	Is there evidence for a collisionally-induced propagating density wave in NGC 253? 
GALEX UV images are examined in an effort to answer this question. The NUV and FUV 
morphology of NGC 253, corrected to show the galaxy as it would appear if viewed face-on,
is shown in Figure 19. The de-projection involved stretching 
the image along the minor axis to simulate tilting the line of sight, and then applying 
a deconvolution filter to correct for distortions introduced by this process. 
The overall distribution of star-forming regions is not affected by the deconvolution process.
No attempt was made to correct for the inherent thickness of the NGC 253 disk, and emission 
from the outflow blurs the right and left hand edges of the disk in the de-projected images. 
The position angle and ellipticity measured from 2MASS images (Jarrett et al. 
2003) were adopted, and residual oblateness in the de-projected images suggests that 
the ellipticity was underestimated by a small amount. 
The side lobes that bracket bright point sources are artifacts of the deconvolution process. 

	The de-projected $K$-band 2MASS image of NGC 253 is also shown in Figure 19. This 
is an interesting comparison dataset as near-infrared light traces mass, whereas UV light 
mainly traces young stars; hence, differences between the near-infrared and UV images are
expected. The barred morphology of the central regions of the galaxy is clearly 
evident in the $K$ image, as are the two spiral arms that emerge from the ends of the bar. 
The high dust content of NGC 253 almost certainly affects the appearance 
of the galaxy in the UV, and the bar that is prominent in the 
near-infrared is invisible in the UV. Dust in the outflow likely produces 
the streaks in the right half of the southern disk of the UV images 
as well as the lack of UV emission immediately to the left of the nucleus.

	The UV sources loosely follow the spiral structure that is conspicuous in the 
near-infrared. The agreement between the near-infrared arms and UV sources
in the northern half of the galaxy is reasonable; while there is some UV emission along 
the spiral arm, the brightest sources of UV emission fall along the leading 
edge of the spiral arm, as expected if the leading edge of the density wave compresses 
the ISM and triggers star formation. There is similar agreement between UV emission and 
spiral structure in the southern half of the disk. With the caveat that there is substantial 
extinction throughout the NGC 253 disk, there is no evidence for ring-like distribution 
of star-forming regions. If a galaxy passed through the NGC 253 disk then it did not leave a 
signature like that predicted by the encounter modelled by Block et al. (2006). This might 
indicate that the companion was not as massive as M32 and/or that the passage had an oblique 
collision geometry.

	We close the discussion of the disk by noting that the north east disk of NGC 253 
bisects the extraplanar HI found by Boomsma et al. (2005), suggesting 
a causal relationship between the elevated SFRs and the gas emission. 
The high levels of recent star formation in the north east 
portion of the NGC 253 disk may have an impact on the circumgalactic environment 
through the formation of chimneys, which are conduits for channeling gas out of the 
disk plane (e.g. Norman \& Ikeuchi 1989). Significant mass outflow rates do not require large 
SFRs (e.g. Ceverino \& Klypin 2009).

	Melioli et al. (2009) model the impact on the extraplanar environment 
of SNe from a star-forming complex in a Milky Way-like system. Their simulations 
indicate that the moderately compact concentration of star-forming regions produces a 
vertical plume of gas that diffuses only modest distances 
from the chimney axis after one disk rotation ($\sim 
200$ Myr in their model). Gas is ejected $\sim 5$ kpc above the disk plane in these 
models, whereas the extraplanar HI near NGC 253 is traced out to 12 kpc 
(Boomsma et al. 2005). Figure 2 of Boomsma et al. (2005) shows 
extended spurs in the extraplanar HI emission, especially the 
column that extends to the north west of the disk plane, and this 
is consistent with the morphology of the chimney outflow predicted by Melioli 
et al. (2009). While the arguements presented here are not ironclad, feedback 
from a `boiling disk' (Sofue et al. 1994), of which the concentration of RSGs 
in the north east disk are the most conspicuous manifestation at near-infrared wavelengths, 
is a viable mechanism for producing the extraplanar HI in NGC 253.

\subsection{The Spatial Distribution and Origins of the Extraplanar Stars}

	A diffuse extraplanar stellar component, which comparisons with model LFs 
in \S 5 indicate contains stars that formed over Gyr or longer timescales, is seen in 
all three WIRCam fields. The distribution on the 
sky of the extraplanar component appears to be flattened, with an ellipticity near 0.4. 
It can be anticipated that stars belonging to this extraplanar 
component should also be found to the north of the galaxy disk.

	The extraplanar regions of NGC 253 contain clues about the 
events that triggered the starburst. If the starburst was triggered by a tidal 
encounter with another galaxy then detritus from the interaction may lurk in the extraplanar 
regions, and evidence for such material is seen in Figure 3. 
While the extraplanar stars are widespread, girdling at 
least the southern part of the galaxy, there are minor field-to-field differences 
in extraplanar stellar density, indicating that the extraplanar 
stars are not yet well mixed throughout the halo. Smaller scale structures, 
such as streams and shells, may become better characterised with deeper images that 
sample larger numbers of objects, such as stars on the RGB. In a related vein, studies of the 
present-day extraplanar component can also be used to predict the future appearance of 
the NGC 253 `halo'; for example, will an observer in the distant future find that the halo 
of NGC 253 contains a complex mix of stellar components, the origins of which can only 
be identified through extremely deep photometric studies? 

	To the extent that the simulations run by Fardal 
et al. (2007; 2008) can be applied to NGC 253, then any tidal streams may have lower 
surface brightnesses than the morass of stars that surround the NGC 253 disk. 
Selection effects may also come into play, as higher surface 
brightness tidal structures tend to be associated with material accreted from more metal-rich 
companions that have been recently accreted (Gilbert et al. 2009). 
Hence, the prominence of stellar tidal streams provides clues about 
the nature of the companion and the time of the encounter. 

	The extraplanar regions of NGC 253 contain a complex mix of stellar populations. 
NGC 253 is almost certainly surrounded by an old, classical stellar halo, although the 
spatial distribution and specific frequency of globular clusters hints that the classical 
halo may be even more poorly populated than the classical halos around 
other galaxies, and have a highly flattened morphology. 
Comeron et al. (2001) find stars with young ages near the minor axis of NGC 253, that 
presumably are associated with the outflow from the nuclear starburst. 
The first stars that formed in the outflow (i.e. at least a few tens of Myr in the past; 
Pietsch et al. 2000) may have diffused away from their places of birth by now, and 
the dominant gravitation influence of NGC 253 will pull the young stars towards the 
disk plane, where they may ultimately form a more compact, kinematically relaxed 
distribution. A fraction of stars that formed in the outflow may also obtain relatively 
high velocities (e.g. Stone 1991), allowing them to move large distances from their places of 
birth.

	The diffuse extraplanar component contains stars 
that are far too young to belong to a classical halo, while the wide age span ($\geq 1$ 
Gyr) deduced from the M giant LF is not consistent with formation in the 
outflow, which has probably been in place for (at most) only a few 100 Myr (i.e. since 
the interaction with a companion). The uniform mix of the extraplanar population over large 
spatial scales, coupled with a LF that is indicative of an extended star-forming 
history that is typical of disk systems, is consistent with the stars having formed in an 
environment other than the extraplanar regions of NGC 253. While it is argued here that the 
stars did not form in the extraplanar regions, pockets of star formation 
in such environments are not without precedent (e.g. Tullmann et al. 2003), 
and {\it in situ} formation has been suggested as a possible mechanism for the formation of 
young stellar ensembles near M82 (Davidge 2008e). 

	Two possible sources of origin for the diffusely distributed extraplanar stars are 
the disk of NGC 253 or a gas-rich companion galaxy with M$_B < -15$. 
An interaction between NGC 253 and another galaxy could 
perturb the orbits of stars in the NGC 253 disk, and produce an extraplanar component.
As for an origin in a dwarf galaxy, NGC 253 does 
not have an entourage of companions, and so the 
companion in question must have been completely destroyed or absorbed by NGC 253.

	The chemical composition of stars in the extraplanar regions of NGC 253 will provide 
obvious clues into their origins. If these stars have near-solar metallicities than this 
would be indicative of having formed in the disk, whereas a metallicity that is, for example, 
one-tenth solar would favor formation in a lower luminosity dwarf galaxy.
The photometric properties of RGB stars are one means of estimating 
metallicities, albeit for stars that formed at least a few Gyr in the past. 
The deep CMD discussed by Mouchine (2006) suggests that the majority of stars in the 
extraplanar regions near the minor axis have a mean [M/H] $\sim -0.7$, ostensibly suggesting 
an origin in a dwarf companion. However, this mean metallicity assumes that the stars are 
all `old'. If the extraplanar regions are populated by stars that formed over a 
range of ages, then a large fraction may have ages $< 3$ Gyr, and the metallicities of 
these objects will be underestimated if they are assumed to be old.

	The combination of visible and near-infrared data would provide considerable 
wavelength leverage for measuring improved metallicities. Still, the ability to measure the 
characteristic metallicity of RGB stars may be complicated by 
the extensive dust content of the galaxy. Differential reddening 
causes conspicuous blurring of the CMDs of NGC 253 in the disk (e.g. the discussion of the 
Karachentsev et al. 2003 and Dalcanton et al. 2009 CMDs in \S 1 of this paper). There is also 
evidence for substantial quantities of dust in the extraplanar regions of the galaxy (Kaneda 
et al. 2009), which probably was ejected by winds from star-forming regions. The broadening of 
sequences on the CMDs of NGC 253 fields caused by differential extinction may be mistaken 
for a mixture of ages and/or metallicities.

	Moderate resolution specroscopy provides a means of measuring metallicities 
that is less susceptible to differential extinction. Given the low 
temperatures of the brightest stars in the extraplanar regions of 
NGC 253, coupled with the need for high angular resolutions to resolve sources 
at small offsets from the disk plane, near-infrared spectra 
are of obvious interest, and the feasibility of obtaining such data
is briefly discussed here. The near-infrared spectral region contains a number of 
metallicity-sensitive features that can be examined with spectral resolutions of $\sim 1500$, 
and the calibration of these has been investigated 
by Frogel et al. (2001). The brightest M giants in 
the extraplanar regions of NGC 253 have $K \sim 19$, and experience with 
spectrographs on 8 -- 10 meter telescopes indicates that total 
on-source exposure times of hundreds of hours would be required to obtain a spectrum 
with a S/N = 100 of a $K = 19$ source with a resolution $\sim 10^3$. However, the exposure 
time is markedly lower if the observations are conducted with a 30 meter telescope 
working at the diffraction limit, due to the D$^4$ advantage. 
Indeed, a simple scaling of the exposure times required for 8 meter telescopes suggests that 
spectra with a suitable S/N ratio could be obtained with exposure times of 
only a few hours with a 30 metre telescope. Thus, while 
obtaining spectra of the brightest red stars is not feasible with 8 -- 10 metre 
telescopes, this is a project that is well within the grasp of a 30 meter telescope.

	NGC 247 is the nearest large galaxy to NGC 253. 
Could NGC 253 and NGC 247 have interacted in the not too distant past?
An interaction between NGC 247 and NGC 253 seems unlikely at first blush, 
as NGC 247 does not show classical signatures of an interaction, such as a central 
starburst or extended tidal arms. Surveys also have not found HI streams near NGC 247 
or NGC 253 (Haynes \& Roberts 1979; Putman et al. 2003) that could be part of a debris trail. 
Still, the disks of these galaxies are oriented in a manner that is consistent with mutual 
tidal torquing (Whiting 1999). In addition, NGC 247 -- like NGC 253 -- has 
a truncated HI disk (Carignan \& Puche 1990), a tendency for mean age to increase 
with radius in the outer disk (Davidge 2006), and -- what is perhaps most suggestive of 
a recent encounter -- an intermediate age extraplanar population (Davidge 2006). 

	If NGC 247 and NGC 253 had a close encounter then tidal fragments may have been 
removed from one or both galaxies. While there are a number of nearby dwarf 
irregular and dwarf spheroidal galaxies, these are located to the north west of 
NGC 253 and are clustered near NGC 247 (e.g. Figure 3 of Karachentsev et al. 2003). 
Could some of the actively star-forming dwarf galaxies near NGC 247 be tidal dwarfs? 

	Tidal dwarfs have characteristics that are distinct from those of classical 
galaxies that formed within primordial dark matter halos. 
The gas and stars that tidal forces displace from a galaxy quickly develop 
different kinematic properties, as gas is a collisional 
system that can dissipate energy, while stellar ensembles are collisionless systems. 
Gas may thus form concentrated pockets where star formation may subsequently occur. 
Because tidal dwarfs are not in dark matter dominated halos, they have 
structural properties that are distinct from those of classical galaxies, and may 
dissipate on timescales that are significantly less than the Hubbble time. 
In addition, they will not have an underlying concentrated substrate of 
older stars, of which RGB stars are among the most prominent signature. 
Finally, if the tidal gas came from a massive galaxy then the stars in a tidal 
dwarf will have a metallicity that differs from that expected from the [M/H] -- 
M$_B$ relation defined by classical galaxies.

	There are two galaxies close to NGC 247 on the sky that have conspicuous blue 
populations: DDO 6 and DDO 226. Karachentsev et al. (2003) present $(I, V-I)$ CMDs 
of both galaxies that are based on moderately-deep WFPC2 observations. 
The CMD of DDO 6 has a prominent RGB, indicating that it contains stars with 
ages of at least a few Gyr. A subsequent investigation of the star-forming history of DDO 6 by 
Weisz et al. (2006) suggests that the majority of stars in DDO 6 formed 12 -- 14 Gyr in the 
past, and that star formation was only rejuventated within the past 0.5 Gyr. Thus, 
based on the presence of a large old population, DDO 6 is not a tidal dwarf.

	The situation might be different for DDO 226, the CMD of which lacks a prominent RGB. 
There {\it is} a concentration of red stars in the CMD with $I \geq 24$, but these 
do not form a well-defined sequence. Working on the assumption that the red stars are on 
the RGB, Karachentsev et al. (2003) assign DDO 226 a distance 
modulus of 28.5, placing it $\sim 1$ Mpc beyond NGC 247 and NGC 253. Deeper 
images will help establish the nature of DDO 226. More specifically, the detection 
of a well-defined RGB would indicate that DDO 6 is not a tidal fragment. 

\subsection{NGC 253 and M31: Different Galaxies with Similar Genealogies?}

	The manner and timescale in which a galaxy and its companions evolve depends on 
initial conditions, such as the orbital characteristics of satellites, and 
environment, including the distances to the nearest large galaxies. While no two galaxies have 
the same initial conditions, and hence do not have the 
same evolutionary histories, the study of resolved stars can 
be used to identify nearby galaxies that may have experienced similar events, 
but are in different stages of their evolution and as such may appear to be unrelated. 
Identifying such relationships will lead to a better understanding 
of galaxy evolution, as it will then be possible to directly probe 
the influence of -- for example -- local environment. 

	M31 is potentially an interesting comparison object for NGC 253. Not only is M31 
sufficiently close that it can be studied in considerable detail, but NGC 253 and 
M31 have similar M$_B$ and M$_K$. The distribution of stars outside of the disk plane in 
M31 (e.g. Figure 1 of Fardal et al. 2007) is also reminiscent 
of that seen in NGC 253. Still, the validity of pairing M31 and 
NGC 253 may seem dubious, given the obvious differences in overall morphology and environment. 
Indeed, NGC 253 lacks the large bulge that dominates the central regions of M31, while M31 is 
not presently experiencing a starburst.

	NGC 253 also lacks an entourage of companions, whereas M31 has many accompanying 
galaxies. The extraplanar structure around M31 indicates that it has had an extremely eventful 
past, and the number of companions in the past was probably even greater than at present. 
Evidence for this comes in the form of conspicuous streams and shells, which are the 
remnants of one or more satellites that have all but disintegrated (e.g. Fardal et al. 2007, 
2008; Ibata et al. 2007; Koch et al. 2008); Tanaka et al. (2010) distinguish 16 substructures 
in the outer regions of M31. The globular cluster system of M31 is also highly 
inhomogeneous, suggesting a diversity in origins (Beasley et al. 2005).

	In addition to the well-documented structures in the outer regions of 
M31, there is also evidence of a diffuse extraplanar stellar component. 
Some of these stars belong to a classical metal-poor halo 
(e.g. Chapman et al. 2006; Kalirai et al. 2006; Ibata et al. 2007; Koch et al. 2008). 
However, the majority of stars in all but the outermost fields have characteristic 
ages and metallicities that differ substantially from those associated with a 
classical halo (e.g. Bellazzini et al. 2003; Brown et al. 2007; Richardson et al. 2009; 
Tanaka et al. 2010).

	In addition to classical halo stars and tidal debris, Brown et al. 
(2007) argue that the extraplanar regions of M31 also
contain stars that formed in the disk of M31, and were displaced during intermediate 
epochs in a merger event. Evidence has been presented in this paper that a similar event 
may have happened in NGC 253, probably within the past few 100 Myr. If NGC 253 
evolves in isolation in the future and does not experience another 
interaction, which seems likely given its isolation, then 
the diffuse extraplanar component will evolve passively, slowly fading and 
becoming less conspicuous; the differences between the old populations that have been 
in place since the initial formation of NGC 253 and the diffuse extraplanar component that 
formed recently will thus become less obvious, as the main sequence turn-off of the most 
recently formed stars dims. The end result will be a mix of old and intermediate age stars 
smilar to what is seen in the outer regions of M31 today. 
Despite the obvious differences in the evolution of NGC 
253 and M31, the present day NGC 253 may provide a glimpse of a younger M31.

\acknowledgements{Sincere thanks are extended to the anonymous referee for providing a 
comprehensive and thoughtful review that lead to a greatly improved paper.}

\appendix

\section{The Detection of a Population of Extraplanar Main Sequence Stars With Ages $\leq 1$ Gyr in the UV} 

	The WIRcam observations have revealed extraplanar AGB stars with ages of 
a few hundred Myr throughout the southern half of NGC 253. There should then be a 
diffuse population of main sequence turn-off (MSTO) stars with masses $\sim 2 - 3$ 
M$_{\odot}$. While such blue stars are too faint to be detected with WIRCam, 
they are relatively bright at near-ultraviolet (NUV) and far-ultraviolet (FUV) wavelengths, 
and if they are present in sufficient numbers then they will produce a diffuse UV 
halo surrounding NGC 253. Given that hotter, younger main sequence 
stars should be confined to the thin disk of NGC 253, then a 
UV color gradient off of the disk plane might also be expected. 

	Archival GALEX NUV and FUV images of NGC 253, which are 
those described by Hoopes et al. (2005), were used 
to search for emission from a diffuse population of moderately bright MSTO stars. 
A $\pm 5.5$ kpc wide region on either side of the minor axis of NGC 253 was excluded from 
the analysis, to avoid confusion with sources of emission associated with the outflow. 
To facilitate the analysis, the UV emission from intermediate age stars was assumed to be 
axially symmetric about both the major and minor axes. This assumption boosts 
the S/N ratio, and is reasonable given that bright extraplanar 
AGB stars permeate the WIRCam fields, and so likely surround the galaxy. 

	NUV and FUV light profiles were obtained by averaging the data in the direction 
parallel to the disk plane at distances between 5.5 and 14 kpc from the minor axis.
NUV and FUV light above the background level is detected out to $z_{dp} = 15$ kpc, 
in good agreement with the spatial distribution of M giants in Field 3. In addition, while 
the FUV -- NUV color is constant between 0 and 10 kpc from the major 
axis, it becomes $\sim 0.6$ magnitudes redder between $z_{dp} = 10$ and 15 kpc, 
signalling a change in the SED of the dominant sources of emission.

	The GALEX data thus indicate that there is diffuse extraplanar UV 
emission around NGC 253, as predicted from the star counts at near-infrared wavelengths.
This emission is not associated with the outflow (e.g. light from the 
nucleus that is scattered by outflowing dust), as a $\pm 5.5$ kpc 
`exclusion zone' was employed on either side of the minor axis when examining the GALEX data. 
The spatial extent of the UV emission off of the disk plane is consistent with that determined 
from resolved AGB stars. That the FUV -- NUV color within 10 kpc of the disk 
plane is constant is consistent with a disk-like stellar mix, again in agreement with 
the study of resolved stars, which indicate that 
the extraplanar AGB LF is similar to that of disk stars.

	The WIRCam and GALEX data allow predictions to be made about the extraplanar 
regions of NGC 253 that can be tested with deeper observations. 
Imaging studies of the extraplanar regions of NGC 253 
at visible wavelengths should reveal a population of blue 
MSTO stars with $V \sim 27 - 28$, that collectively are the source of the diffuse 
UV emission. Two predictions can also be made about the distribution of 
these objects. First, they will be seen out to $\sim 15$ kpc above the disk plane, based on 
the detection of bright AGB stars and UV emission out to this same distance.
Second, the UV color profile suggests that the luminosity-weighted age 
changes $\sim 10$ kpc from the major axis, in the sense that a slightly older population 
dominates when $z_{dp} \geq 10$ kpc than at smaller distances. There should then be 
a change in the brightness of the MSTO near $z_{dp} \sim 10$ kpc.

\parindent = 0.0cm

\clearpage

\begin{table*}
\begin{center}
\begin{tabular}{ccc}
\tableline\tableline
Field \# & RA & Dec \\
 & (2000) & (2000) \\
\tableline
1 & 00 46 30.0 & --25 30 00 \\
2 & 00 48 45.0 & --25 05 00 \\
3 & 00 48 30.0 & --25 30 00 \\
\tableline
\end{tabular}
\end{center}
\caption{Field Co-ordinates}
\end{table*}

\clearpage

\begin{figure}
\figurenum{1}
\epsscale{0.75}
\plotone{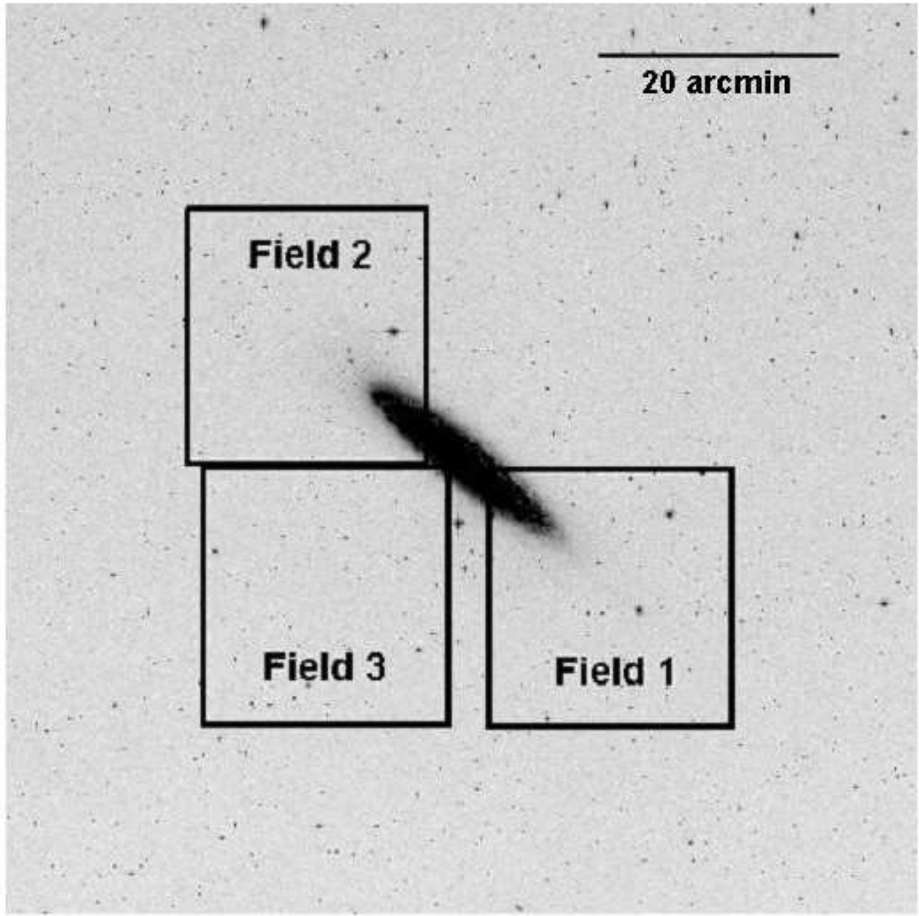}
\caption
{A $1.5 \times 1.5$ degree$^2$ section of the Digital Sky Survey in the blue filter showing 
the locations of the three WIRCam fields. North is at the top, and East is to the left. 
The angular scale is indicated.}
\end{figure}

\clearpage

\begin{figure}
\figurenum{2}
\epsscale{0.75}
\plotone{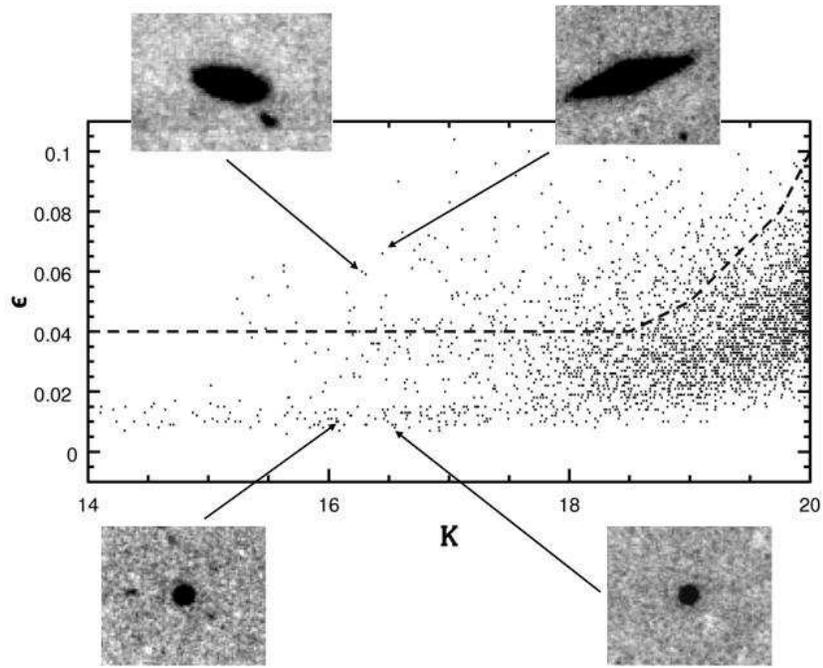}
\caption
{The error computed by DAOPHOT, $\epsilon$, as a function of $K$ magnitude for 
sources in Field 3. The dashed line shows the relation that was used to cull stars from the 
photometric catalogue - objects that fall above this line were rejected from the catalogue. 
Images of objects selected at random to demonstrate the types of objects that are 
associated with different locations on the $\epsilon - K$ plane are also shown. The 
two highlighted objects with relatively high $\epsilon$ are clearly galaxies, while the 
two highlighted objects with smaller $\epsilon$ are point sources.}
\end{figure}

\clearpage

\begin{figure}
\figurenum{3}
\epsscale{1.05}
\plotone{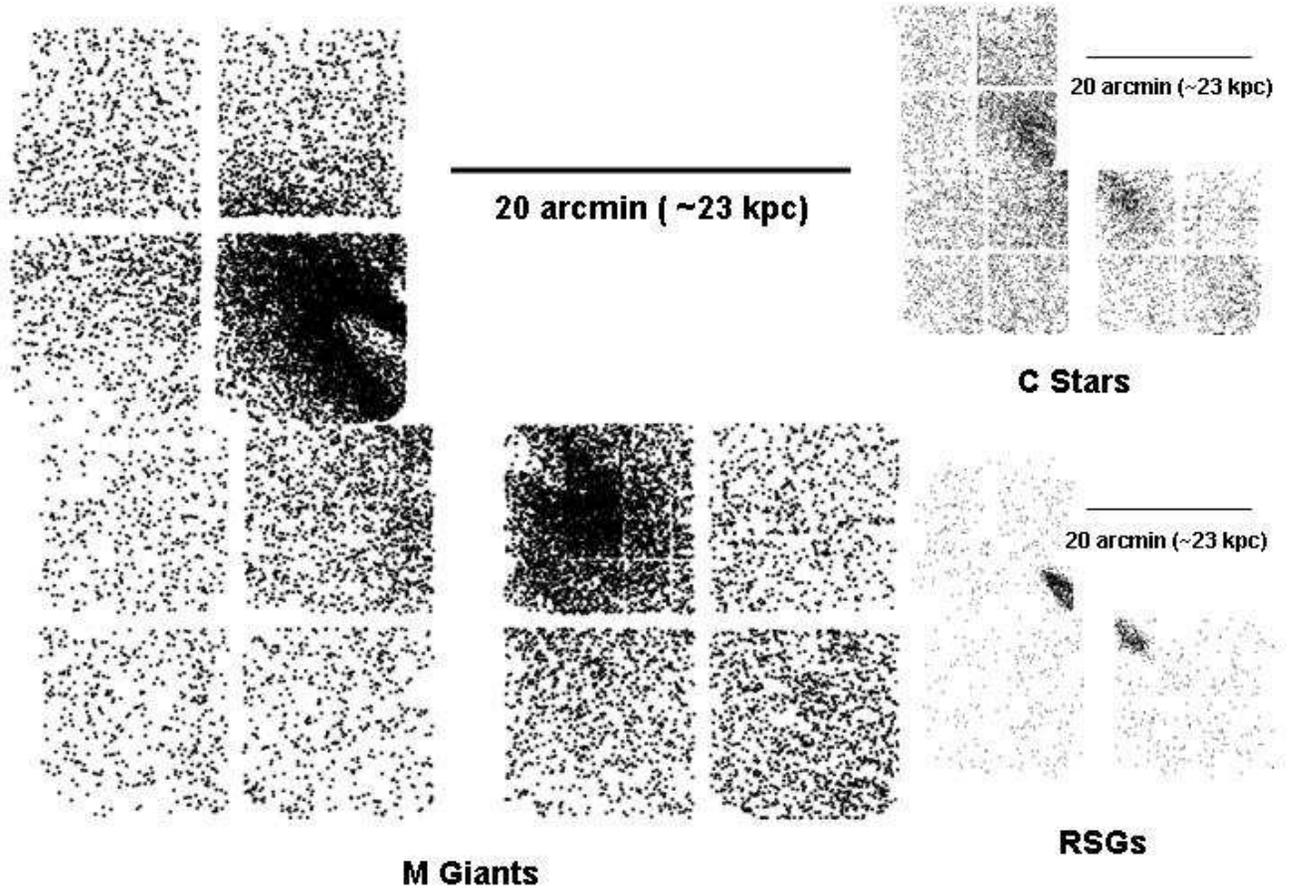}
\caption
{The distribution of M giants, C stars, and RSGs in and around NGC 253. The stars that are 
shown here were selected based on location on the $(K, J-K)$ CMD, as indicated in the 
right hand panel of Figure 6. North is at the top, and East is to the left. 
The angular scale is indicated in the upper right hand corner of each panel, with a 
corresponding projected distance that assumes the Karachentsev et al. 
(2003) distance modulus. The nucleus of NGC 253 is closer to Field 2 than to Field 1. 
Note the lop-sided distribution of M giants, and the diffuse distribution of M giants in 
the region around the disk, with sub-structure close to the disk plane. 
Both of these findings are suggestive of an interaction (or interactions) with 
another galaxy (or galaxies) during intermediate epochs.}
\end{figure}

\clearpage

\begin{figure}
\figurenum{4}
\epsscale{0.75}
\plotone{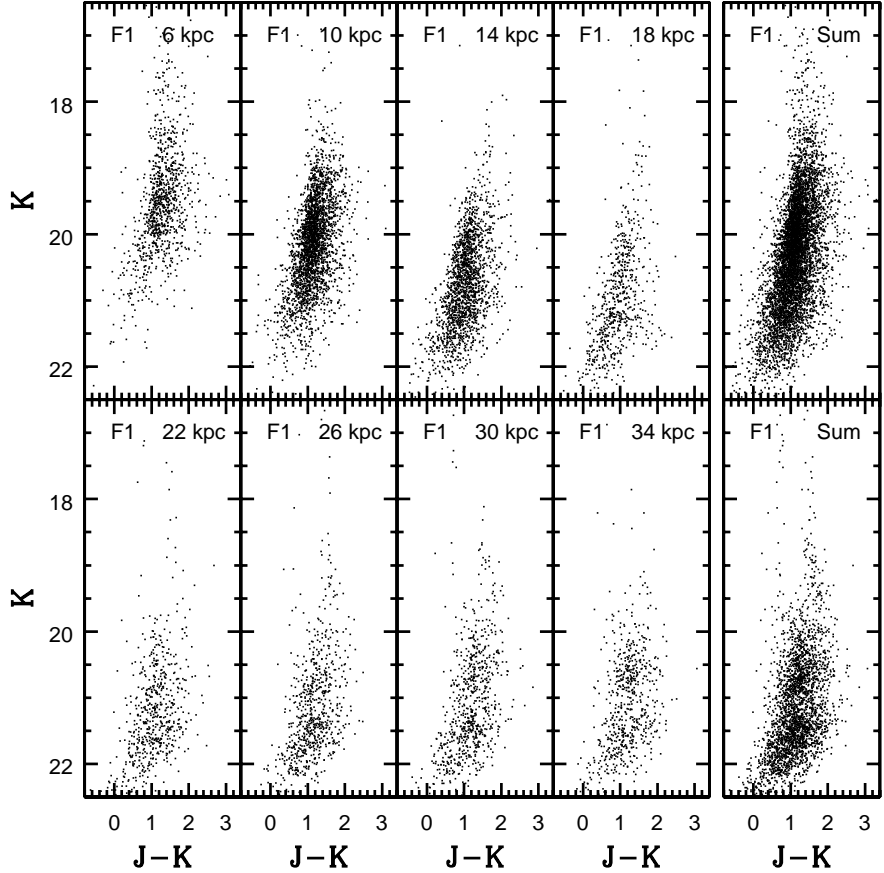}
\caption
{The $(K, J-K)$ CMDs of disk stars in Field 1. Each CMD contains sources
in a 4 kpc wide annulus, extracted assuming that the disk is inclined to the 
line of sight by 76.1 degrees. The distance listed in each panel refers to the 
midpoint of that annulus, as measured along the major axis. 
The righthand panel in each row shows the composite of all CMDs in that 
row, which are shown to illustrate the differences between the sources 
present at R$_{GC} \leq 18$ kpc and R$_{GC} \geq 22$ kpc. The prominent concentration 
of objects in the R$_{GC} \leq 18$ kpc CMDs near $J-K \sim 1.2$ is made up of 
oxygen-rich AGB stars. At the largest R$_{GC}$ the CMDs consist almost exclusively 
of foreground Galactic stars and background galaxies; foreground stars populate 
the vertical finger of sources with $J-K \sim 0.8$, while background galaxies make 
up the diffuse cloud of objects centered near $J-K \sim 1.6$.}
\end{figure}
 
\clearpage

\begin{figure}
\figurenum{5}
\epsscale{0.75}
\plotone{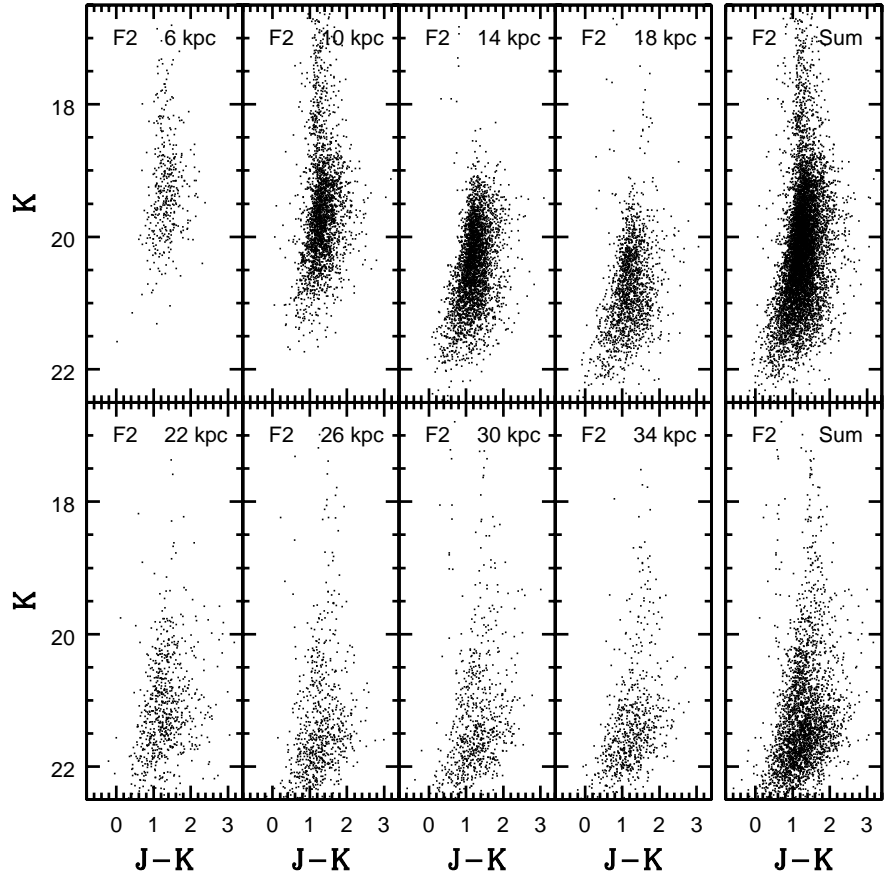}
\caption
{The same as Figure 4, but showing the CMDs of sources in Field 2. Note the prominent 
sequence of RSGs in the R$_{GC} = 10$ kpc CMD. A similar richly populated RSG 
sequence is not seen in the Field 1 CMDs, indicating that Field 2 contains a 
higher fraction of stars that formed during the past $\sim 0.1$ Gyr than Field 1.}
\end{figure}

\clearpage

\begin{figure}
\figurenum{6}
\epsscale{0.75}
\plotone{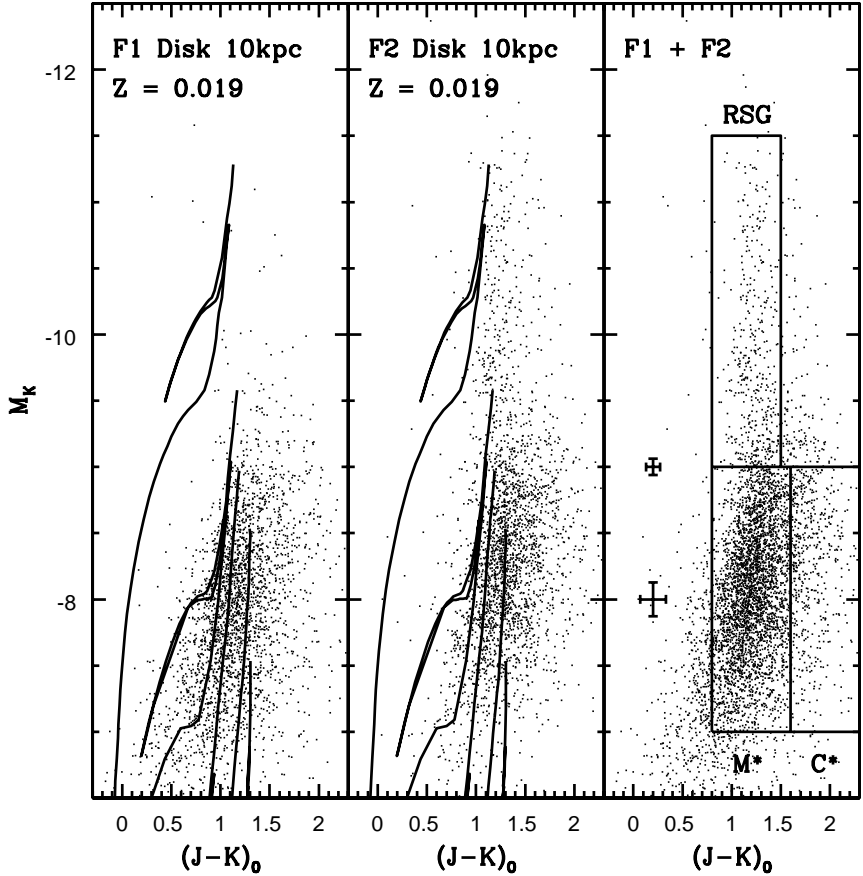}
\caption
{The R$_{GC} = 10$ kpc CMDs of disk stars in Fields 1 and 2 are compared with 
solar metallicity isochrones from Girardi et al. (2002). The error bars in the 
right hand panel show $1\sigma$ uncertainties in the photometric measurements 
as estimated from artificial star experiments - the uncertainties for stars with 
M$_K < -9$ are too small to be displayed at this scale. Isochrones with ages 
10 Myr, 30 Myr, 100 Myr, 1 Gyr, and 10 Gyr are also shown. 
The peak M$_K$ magnitude of RSGs in the Field 2 CMD 
is comparable to the tip of the 10 Myr isochrone, as expected if RSGs do not form in 
systems that are much younger than 10 Myr. The photometry has been corrected 
only for foreground extinction, and the difference in $J-K$ between the isochrones and 
RSGs in Field 2 is probably due to extinction that is internal to NGC 253 (see text).
The majority of the AGB stars detected with WIRCam in NGC 253 have ages of only a few Gyr. 
The areas that are used to define the M giant, C star, and RSG samples that are plotted 
in Figure 3 are indicated in the right hand panel, which shows a composite of the 
10 kpc Field 1 and Field 2 CMDs.}
\end{figure}

\clearpage

\begin{figure}
\figurenum{7}
\epsscale{0.75}
\plotone{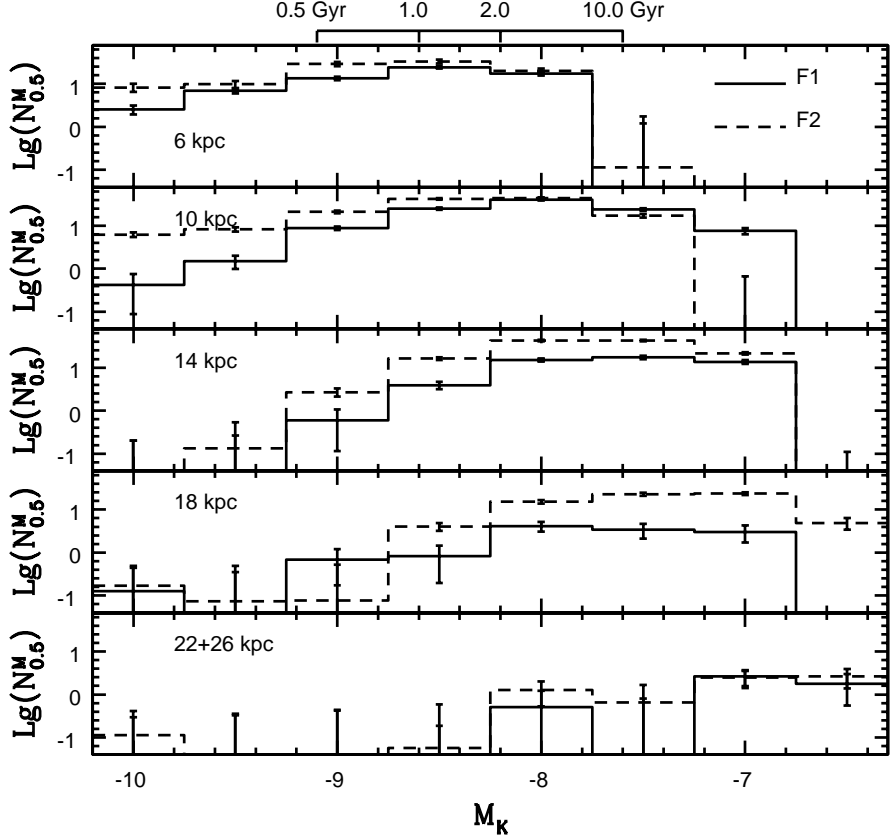}
\caption
{The $K$ LFs of objects with $J-K$ between 0.8 and 1.6, which are M giants (M$_K > -9$) and 
RSGs (M$_K < -9$), in the disk plane of Fields 1 (solid lines) and 2 (dashed lines). 
N$_{0.5}^M$ is the number of M giants per 0.5 $K$ magnitude interval arcmin$^{-2}$ in each 
annulus. Source counts in the R$_{GC} = 30$ and 34 kpc intervals have been 
subtracted from the LFs at smaller radii to correct statistically for foreground stars and 
background galaxies. The error bars show counting statistics. The Field 1 and Field 2 LFs 
differ at the bright end of the 10 kpc LF, owing to the large concentration of RSGs 
in this part of Field 2. The stellar density in Field 2 also exceeds that in Field 1 in 
the R$_{GC}  = 14$ and 18 kpc LFs. The stellar distribution in the outer disk of NGC 253 is 
thus not azimuthally symmetric, but is skewed towards higher stellar densities in Field 2.} 
\end{figure}

\clearpage

\begin{figure}
\figurenum{8}
\epsscale{0.75}
\plotone{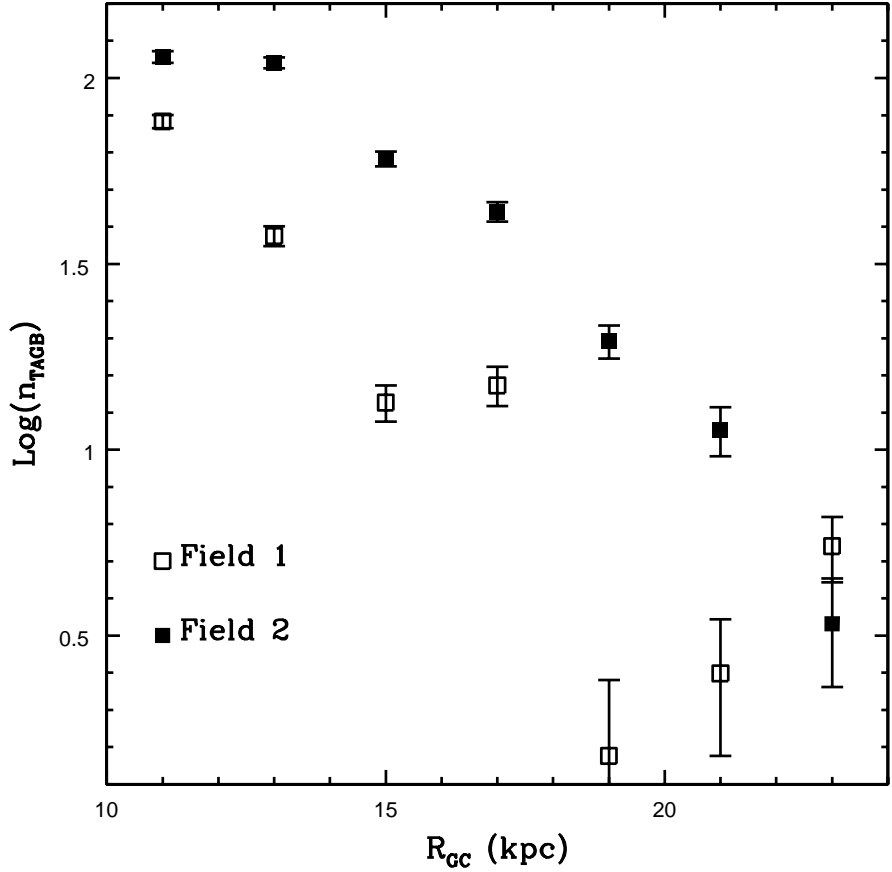}
\caption
{The spatial distribution of the brightest M giants in the 
disk plane of Fields 1 and 2. n$_{TAGB}$ is the number 
of M giants arcmin$^{-2}$ with $K$ between 19.5 and 20.5 and $J-K$ between 0.8 and 1.6, 
corrected for contamination from foreground stars and background galaxies based on 
source counts at large distances from the galaxy center. R$_{GC}$ is distance 
measured in the plane of the disk, assuming a disk inclination 
of 76.1 degrees. The errorbars show the $1\sigma$ statistical uncertainties in 
the source counts. Note that the number counts in Field 2 consistently exceed 
those in Field 1 at the same R$_{GC}$. Also, whereas the Field 2 data follow a well-defined 
exponential profile out to R$_{GC} = 23$ kpc, this is not the case for the Field 1 
points, which show a prominent plateau in number counts near R$_{GC} = 15 - 17$ kpc.}
\end{figure}

\clearpage

\begin{figure}
\figurenum{9}
\epsscale{0.75}
\plotone{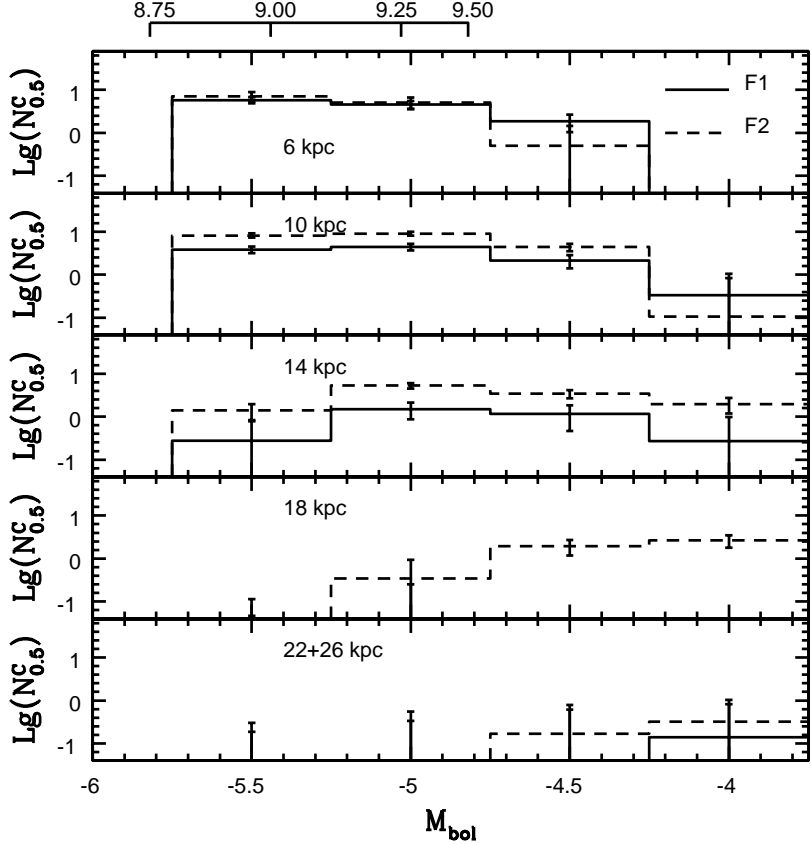}
\caption
{The M$_{bol}$ LFs of objects with $J-K$ between 1.6 and 2.5, which are cool C stars, in the 
disk plane of Fields 1 (solid lines) and 2 (dashed lines). M$_{bol}$ was calculated 
using the Galactic and LMC bolometric corrections from Bessell \& Wood (1984). 
N$_{0.5}^C$ is the number of C stars per 0.5 M$_{bol}$ 
magnitude interval arcmin$^{-2}$ in each annulus. The LFs have been 
corrected for contamination from foreground stars and background galaxies by 
subtracting the number counts of sources in the R$_{GC} = 30$ and 34 kpc annuli. 
The relation between peak AGB M$_{bol}$ and age from the solar metallicity Girardi 
et al. (2002) isochrones is given at the top of the figure. The Field 1 and Field 2 LFs are 
identical at R$_{GC} = 6$ kpc, but become progressively more different as 
R$_{GC}$ increases. Significant numbers of C stars are not seen in Field 1 
at R$_{GC} \geq 14$ kpc, hinting that the C/M ratio changes 
in this part of NGC 253, in the sense of becoming smaller towards larger radii.}
\end{figure}

\clearpage

\begin{figure}
\figurenum{10}
\epsscale{0.75}
\plotone{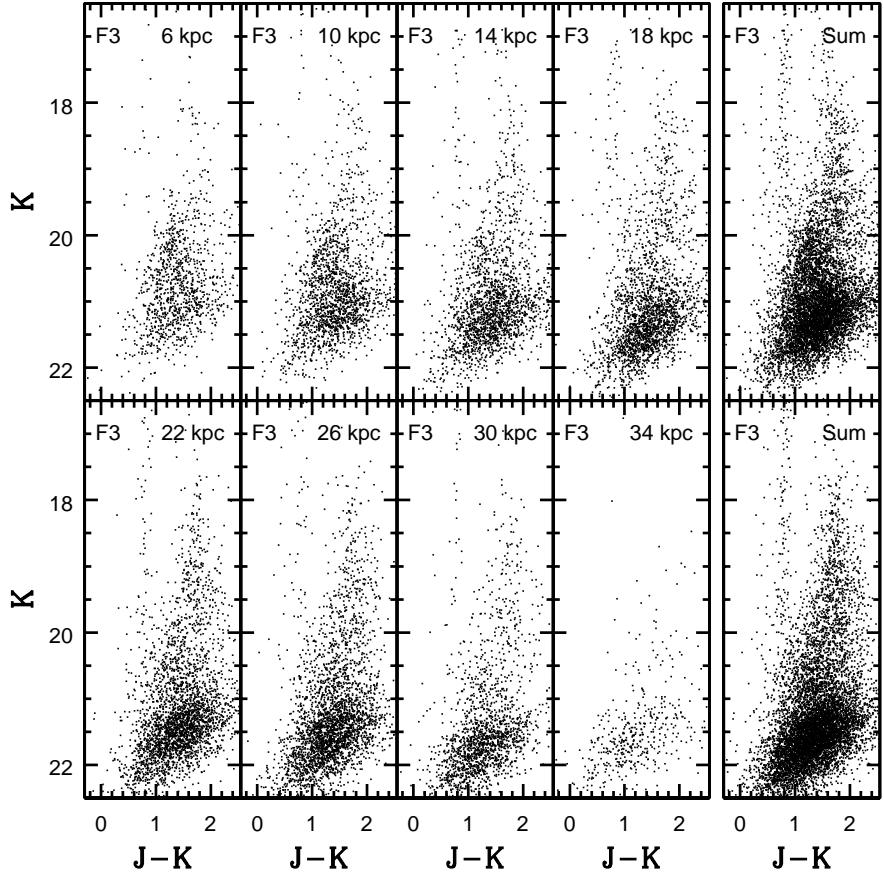}
\caption
{The $(K, J-K)$ CMDs of sources in Field 3, grouped in 4 kpc wide annuli. The distance, 
$z_{dp}$, listed in each panel is measured from the galaxy center on the plane of the sky. 
The panels labelled `Sum' show the composite CMD of all sources in the CMDs in that 
row; these are shown to illustrate the differences between the sources 
present at z$_{dp} \leq 18$ kpc and z$_{dp} \geq 22$ kpc. 
Oxygen-rich AGB stars form a sequence with $J-K \sim 1.2$ and 
$K \geq 20$ that is evident in the CMDs with $z_{dp} \leq 14$ kpc. 
The number of sources per arcmin$^{2}$ remains constant 
when $z_{dp} > 26$, as expected if the photometric catalogue at these 
$z_{dp}$ is dominated by foreground Galactic stars and background galaxies.
While not clearly evident in a visual inspection of the CMDs, a modest number 
of objects that have projected number densities that exceed those 
estimated for foreground stars and background galaxies at the largest R$_{GC}$, 
are detected out to R$_{GC} \sim 22 - 26$ kpc (\S 5.3), 
and these are probably AGB stars that belong to NGC 253.} 
\end{figure}

\clearpage

\begin{figure}
\figurenum{11}
\epsscale{0.75}
\plotone{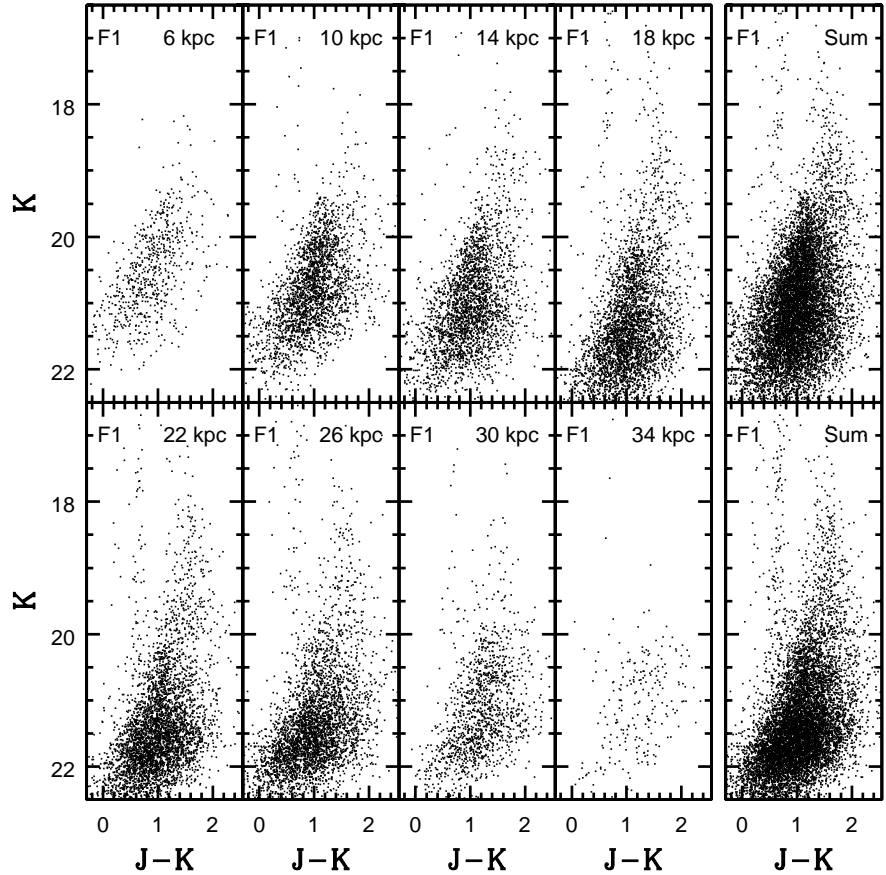}
\caption
{The same as Figure 10, but for extraplanar stars in Field 1. Number counts, corrected 
for foreground stars and background galaxies, indicate that bright AGB stars are present in 
statistically sigificant numbers out to $z_{dp} = 22$ kpc (\S 5.3). Intermediate age 
extraplanar stars are thus not restricted to Field 3, but likely surround the galaxy.}
\end{figure}
 
\clearpage

\begin{figure}
\figurenum{12}
\epsscale{0.75}
\plotone{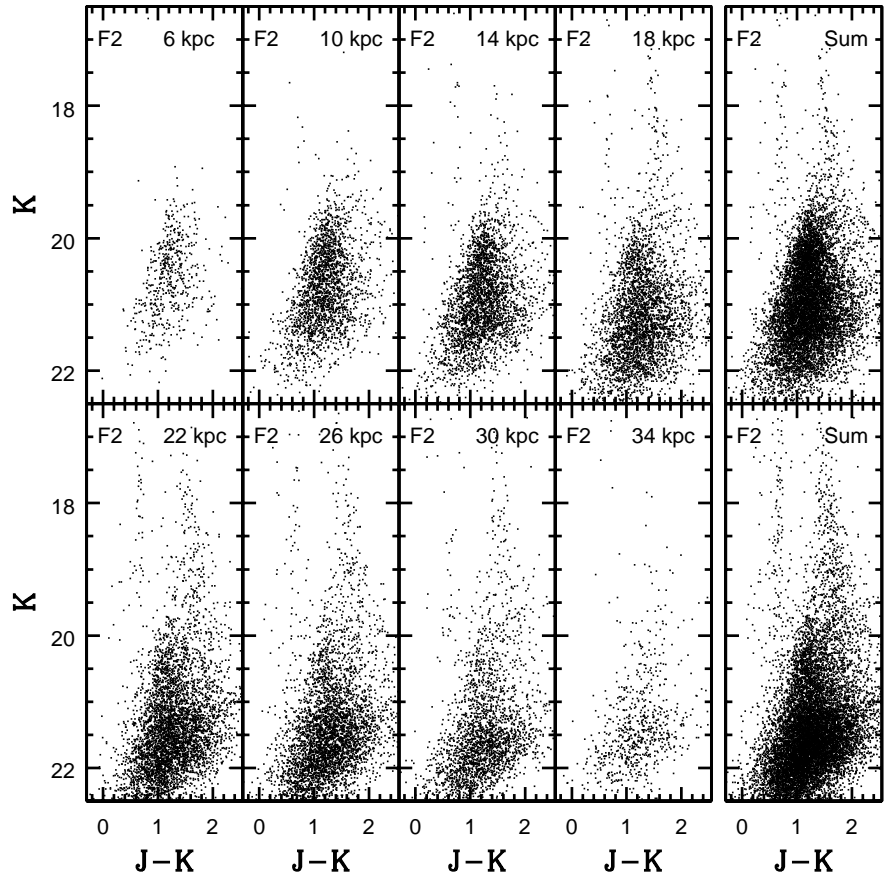}
\caption
{The same as Figure 10, but showing CMDs of the extraplanar regions of Field 2. While 
difficult to identify in the CMDs at large $z_{dp}$ given the contamination from 
foreground stars and background galaxies, number counts indicate that M giant AGB stars are 
present in statistically significant numbers out to $z_{dp} = 22$ kpc (\S 5.3).}
\end{figure}

\clearpage

\begin{figure}
\figurenum{13}
\epsscale{0.75}
\plotone{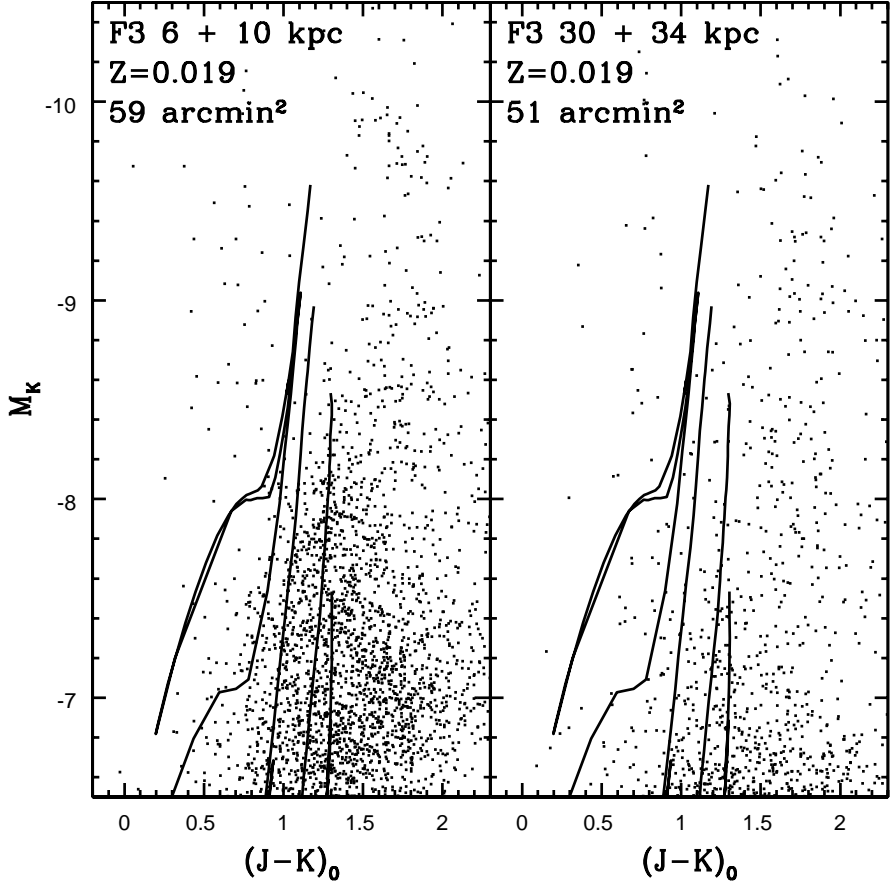}
\caption
{The $(M_K, J-K)$ CMDs of sources in Field 3. 
A distance modulus of 28.0 (Karachensev et al. 2003) and a foreground 
reddening A$_B = 0.05$ mag (Burstein \& Heiles 1982) are assumed. 
The isochrones are from Girardi et al. (2002), with Z = 0.019 
and ages 30 Myr, 100 Myr, 1 Gyr, and 10 Gyr. The total area that is 
sampled in each CMD is also shown. The 30 $+$ 34 kpc CMD 
is dominated by foreground stars and background galaxies. Given that the $6 + 10$ kpc and 
$30 + 34$ kpc intervals sample similar total areas on the sky then 
the $30 + 34$ kpc CMD provides a reasonable sense of the 
number of sources in the $6 + 10$ kpc CMD that do not belong to NGC 
253. There is a clear excess number of objects with $J-K \sim 1 - 1.5$ and M$_K > -8.5$ 
in the $6 + 10$ kpc CMD that is due to AGB stars in NGC 253. The 1 Gyr isochrone passes 
through the middle of this group of objects.}
\end{figure}

\clearpage

\begin{figure}
\figurenum{14}
\epsscale{0.75}
\plotone{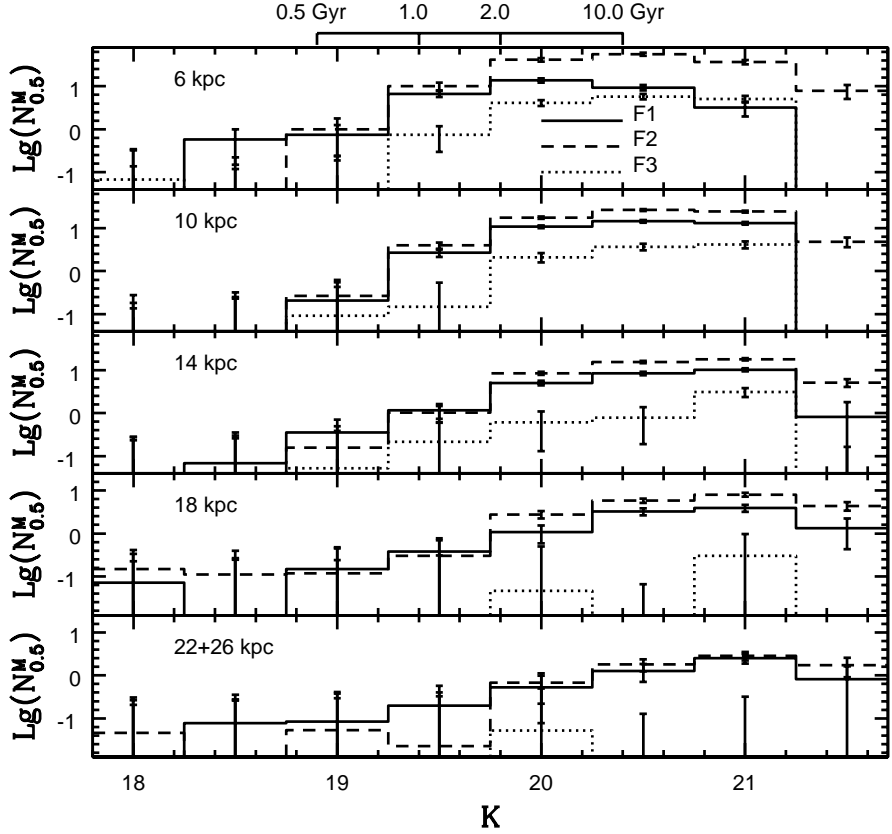}
\caption
{The same as Figure 7, but showing the LFs of stars with $J-K$ between 0.8 and 1.6 
in the extraplanar regions of Fields 1 (solid line), 2 (dashed line), and 3 
(dotted line). The LFs have been corrected for contamination from foreground 
stars and background galaxies by subtracting number counts at large $z_{dp}$. 
Incompleteness causes the number counts to drop at the faint end. 
There is considerable field-to-field dispersion in the number density of sources, with 
the lowest densities in Field 3. This suggests that the extraplanar component 
around NGC 253 is not distributed with circular symmetry on the sky, but follows an 
elliptical distribution with an ellipticity $\approx 0.4$ (\S 5). Despite differences 
in density, the LFs of extraplanar stars in Fields 1, 2, and 3 for $z_{dp} \leq 
14$ kpc have similar shapes, suggesting that the 
extraplanar component around NGC 253 comes from stars that share a common origin; these 
stars are not located in streams or structures that were created at different epochs. 
With the caveat that many highly evolved AGB stars are photometric variables, 
thereby smearing the magnitude of the brightest stars, 
stars with ages $\leq 2$ Gyr appear to be present out to $z_{dp} = 14$ kpc in Field 3, 
and even greater $z_{dp}$ in Fields 1 and 2.} 
\end{figure}

\clearpage

\begin{figure}
\figurenum{15}
\epsscale{0.75}
\plotone{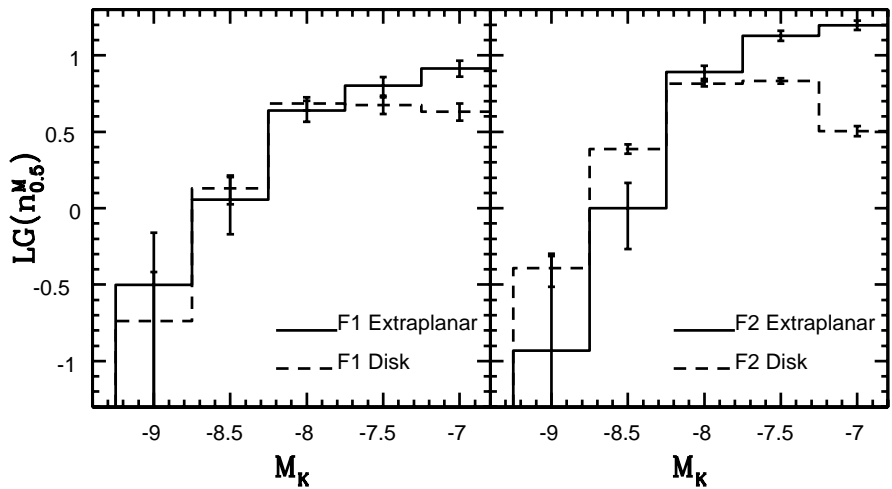}
\caption
{The LFs of extraplanar sources with R$_{GC} = 14$ kpc in Fields 1 (left hand 
column) and 2 (right hand column) are compared with the LFs of sources 
in the disk planes of the same fields. The disk plane LFs have been 
scaled to match the mean counts in the extraplanar LFs at M$_K = -7.5, -8.0,$ and --8.5. 
High stellar density causes incompleteness to become an issue near 
M$_K \sim -8$ in the disk LFs, as opposed to M$_K \sim -7$ in the extraplanar LFs. Still, 
the general agreement between the disk and extraplanar LFs at the bright end suggests 
that the recent star-forming history of the extraplanar component was 
similar to that of the NGC 253 disk.}
\end{figure}

\clearpage

\begin{figure}
\figurenum{16}
\epsscale{0.75}
\plotone{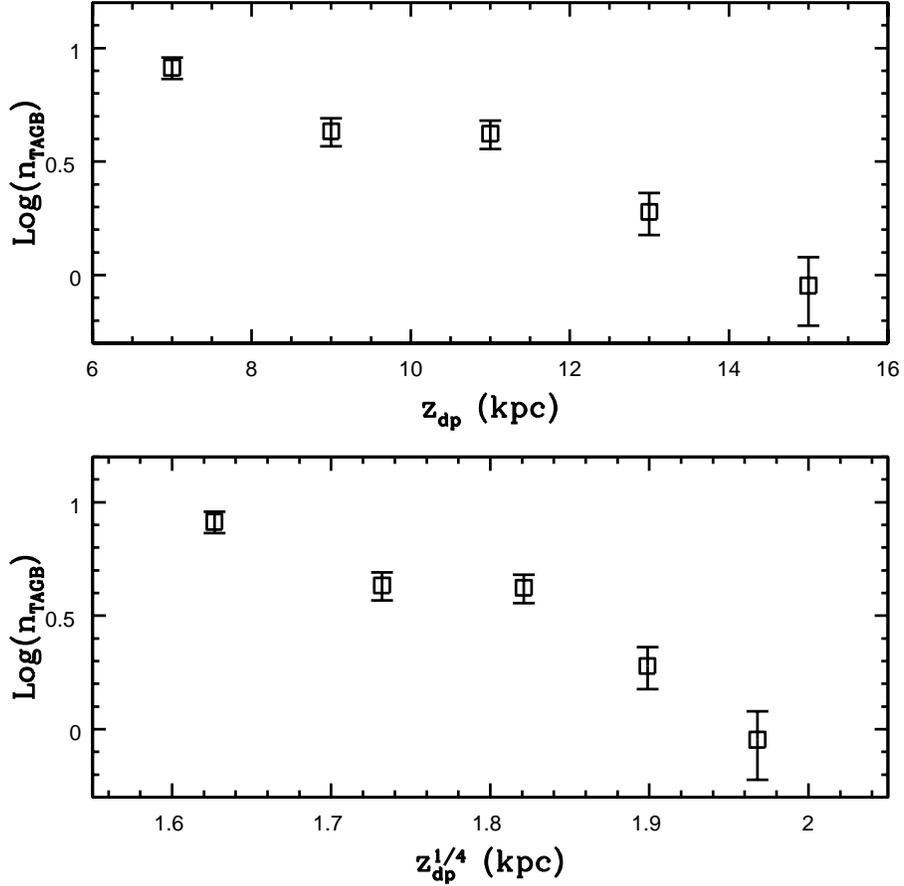}
\caption
{The spatial distribution of the brightest M giants in Field 3. n$_{TAGB}$ is the number 
of M giants arcmin$^{-2}$ with $K$ between 19.5 and 20.5 and $J-K$ between 0.8 and 1.6, 
corrected for contamination from foreground stars and background galaxies using 
source counts near the edge of the field. z$_{dp}$ is the distance from the center of NGC 
253 on the plane of the sky. The errorbars show the $1\sigma$ statistical uncertainties 
in the source counts. The absence of significant discontinuities in the data suggests that 
the brightest M giants in the extraplanar regions of NGC 253 probe a population of objects 
that is well-mixed throughout this part of the galaxy.}
\end{figure}

\clearpage

\begin{figure}
\figurenum{17}
\epsscale{0.75}
\plotone{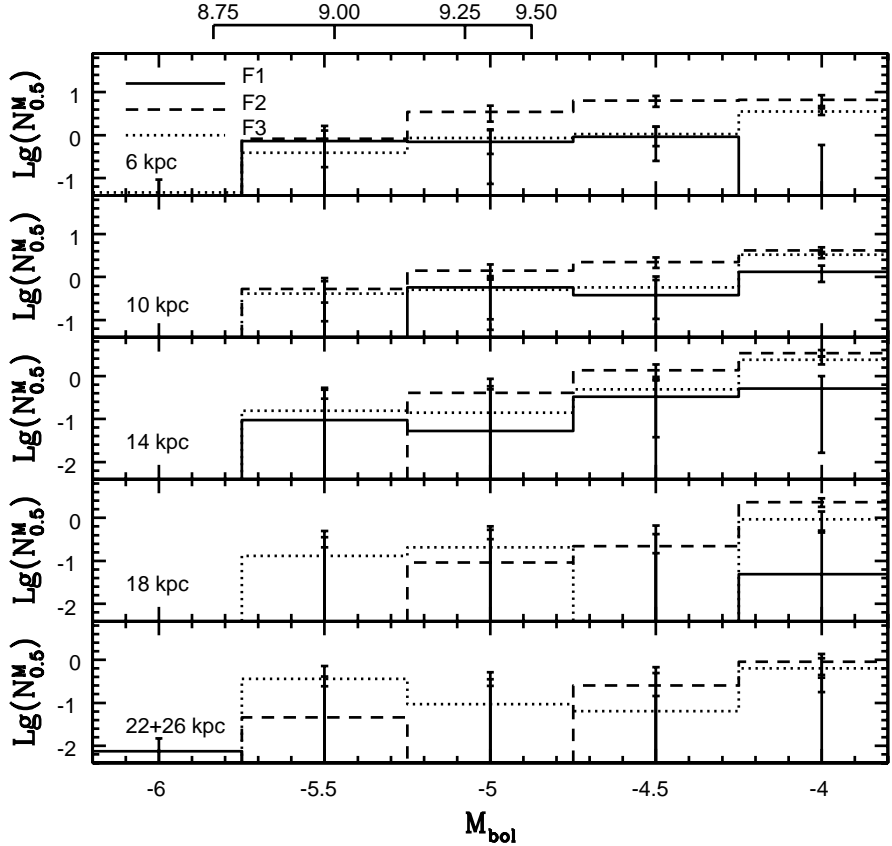}
\caption
{The M$_{bol}$ LFs of extraplanar C stars in Fields 1, 2, and 3. 
N$^{C}_{0.5}$ is the number of C stars per 0.5 M$_{bol}$ magnitude interval arcmin$^{-2}$, 
corrected for foreground dwarfs and background galaxies using the procedure described 
in the text. Significant numbers of C stars are detected out to $z_{dp} = 14$ kpc in all 
three fields, and to $z_{dp} > 20$ kpc in Field 1. The relation between AGB-tip 
M$_{bol}$ and age for solar metallicity stars from the Girardi et al. (2002) isochrones is 
shown at the top of the figure, and this calibration suggests that C stars with ages 
$\leq 1$ Gyr are present throughout much of the extraplanar regions.}
\end{figure}

\clearpage

\begin{figure}
\figurenum{18}
\epsscale{0.75}
\plotone{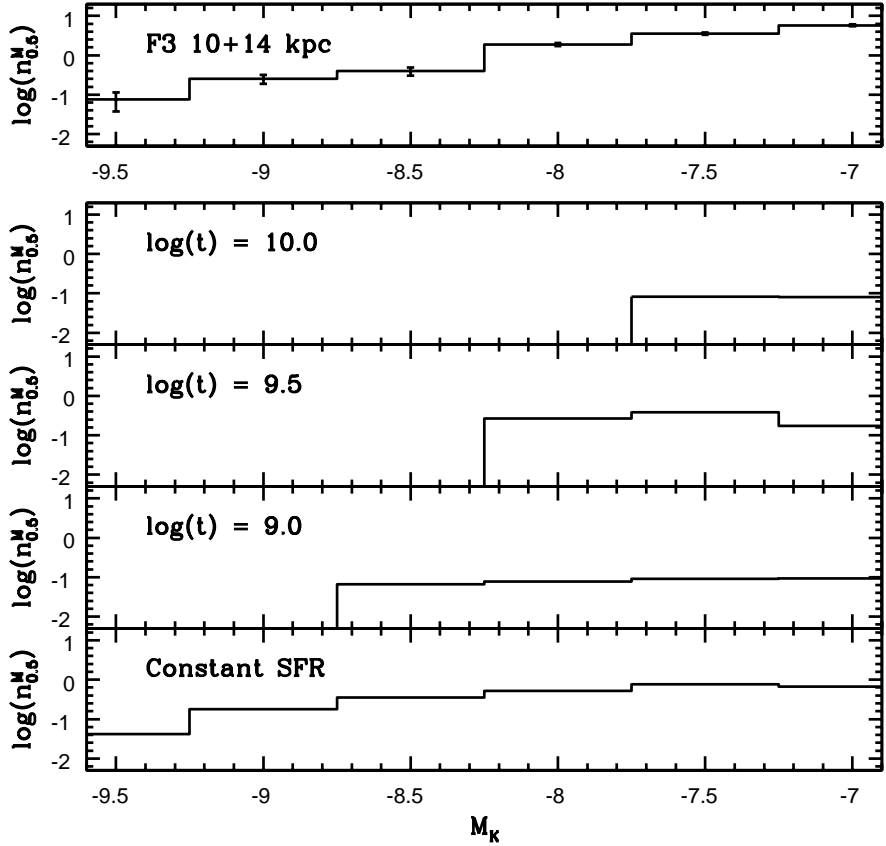}
\caption
{The composite $K$ LF of stars with $J-K$ between 0.8 and 1.6 in the $z_{dp} = 10$ and 14 kpc 
intervals of Field 3 is compared with model LFs that have been constructed from the Z = 
0.019 isochrones of Girardi et al. (2002). n$^{M}_{0.5}$ is the observed number of stars 
per square arcmin per 0.5 magnitude interval, corrected for foreground Galactic stars and 
background galaxies by subtracting source counts at large $z_{dp}$. The models have 
been scaled to fall near the center of each panel for display purposes. The SSP model LFs 
are relatively flat, and the comparatively steep downward slope of the 
observed LF at the bright end indicates that Field 3 contains 
stars that span a range of ages. The panel labelled `Constant 
SFR' shows a model LF for a solar metallicity 
system that has experienced a constant SFR over the time 
interval 0.2 - 10.0 Gyr. While the continuous SFR model better 
matches the downward trend of the observed LFs towards brighter M$_K$ than the SSP LFs, 
the agreement could be further improved by adopting a model in which the SFR decreased 
$\sim 1$ -- 2 Gyr in the past. Nevertheless, the comparisons with the model LFs suggest that 
the intermediate age extraplanar stars in NGC 253 originated in a system that had an 
active star forming history over Gyr timescales.}
\end{figure}

\clearpage

\begin{figure}
\figurenum{19}
\epsscale{1.05}
\plotone{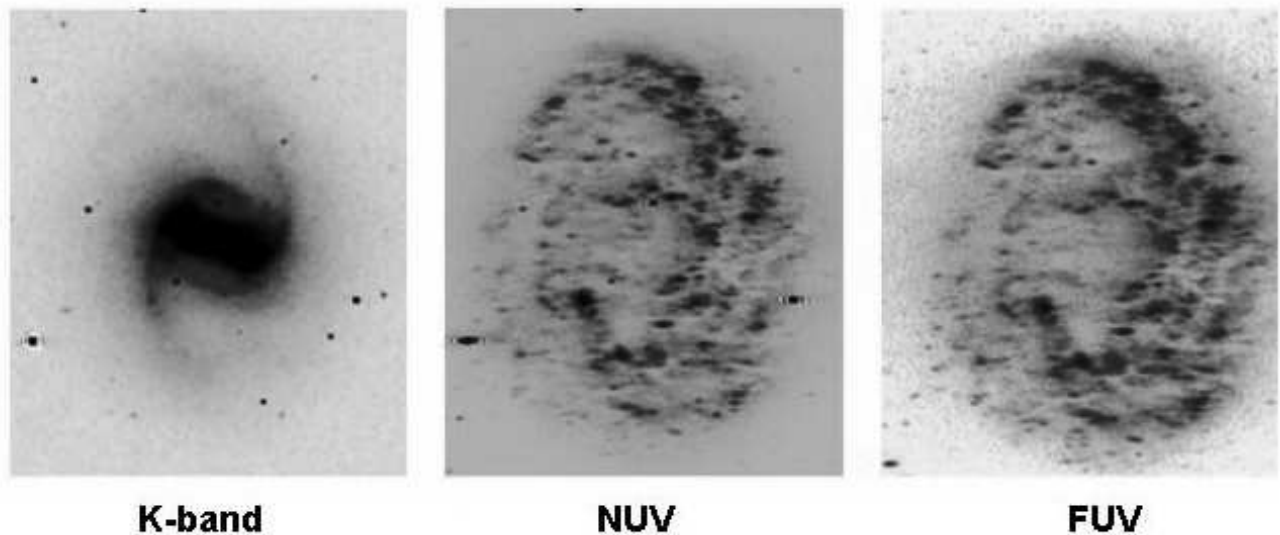}
\caption
{Near-infrared, near UV, and far UV images of NGC 253, processed to show the galaxy as it 
would appear if viewed face-on. The near-infrared image traces total stellar mass, 
while the UV images trace young stars. The images are oriented so that the major axis is 
in the vertical direction. The $K-$band image is from the 2MASS Large Galaxy Atlas (Jarrett 
et al. 2003) while the UV images are the data discussed by Hoopes et al. (2005). 
The sidelobes that bracket bright sources are artifacts of the deconvolution process that was 
applied to correct for distortions introduced during de-projection; the distribution of 
star-forming regions in the de-projected images is not affected by this filtering. 
While the morphology of the galaxy in the UV is undoubtedly affected by dust 
extinction (e.g. the absence of UV light from the central regions), 
there is no evidence for a well-defined star-forming ring, such as might form if a 
moderately massive companion had recently passed near the center of the galaxy.}
\end{figure}

\end{document}